\documentclass{article}

\usepackage{adjustbox} 
\usepackage{arxiv}
\usepackage[utf8]{inputenc} % allow utf-8 input
\usepackage{textcomp} 
\usepackage{hyperref}       % hyperlinks
\usepackage{url}            % simple URL typesetting
\usepackage{booktabs}       % professional-quality tables
\usepackage{nicefrac}       % compact symbols for 1/2, etc.
\usepackage{microtype}      % microtypography
\usepackage{graphicx}
\usepackage{doi}
\usepackage{siunitx}
\usepackage{array}
\usepackage{changepage}
\usepackage{amsmath,amssymb,amsfonts}%
\usepackage{colortbl}
\usepackage{amsthm}%
\usepackage{mathrsfs}%
\usepackage[title]{appendix}%
\usepackage{xcolor}%
\usepackage{textcomp}%
\usepackage{manyfoot}%
\usepackage{algorithm}%
\usepackage{algorithmicx}%
\usepackage{algpseudocode}%
\usepackage{listings}%
\usepackage{subcaption}
\usepackage{svg}
\usepackage{pbox} %for line breaks inside table cells
\usepackage{dsfont}
\usepackage{gensymb}
\usepackage{geometry}
\usepackage{mathtools}
\usepackage{pgfplots}
  \pgfplotsset{
    compat = 1.18,
    axis line style={white},
    every axis label/.append style ={black},
    every tick label/.append style={black},
    every tick/.append style={white}  
  }
\definecolor{mycolor2}{rgb}{1.0,1.0,1.0}%
\usepackage{enumitem}
\usepackage{stfloats} %For big images on same page
\usepackage[
justification=justified]{caption}
\usepackage{placeins}
\usepackage{lineno}
\title{Wave-resolving Voronoi model of Rouse number for sediment entrainment}

\author{ {Johanes Lawen}\\%\thanks{Corresponding author} \\
	\texttt{jl@environment.report} \\
}

\renewcommand{\shorttitle}{\textit{arXiv Physics: Atmospheric and Oceanic Physics}}

\hypersetup{
pdftitle={A template for the arxiv style},
pdfsubject={q-bio.NC, q-bio.QM},
pdfauthor={David S.~Hippocampus},
pdfkeywords={First keyword, Second keyword, More},
}

\begin{document}
\maketitle

\begin{abstract}
To integrate wave and sediment transport modeling, a computationally extensive wave-resolving Voronoi mesh-based simulation has been developed to improve upon heretofore separate sediment and spectral wave modeling. Orbital wave motion-dependent sediment transport and fine structures of the dynamic Rouse number distribution across the seabed were brought into focus. The entirely parallelized wave-resolving hydrodynamic model is demonstrated for nearshore beach waters adjacent to artificial islands in Doha Bay. The nested model was validated with tidal time series for three locations and two seasons.
\end{abstract}

% keywords can be removed

\section{Introduction}\label{sec1}
Wave-resolving hydrodynamics, similar to direct numerical simulations (DNS) of turbulent eddy mixing, requires resolutions that are usually impossible or impracticable to compute within even logarithmic orders of the TFLOP range. Therefore, waves and sediment transport have heretofore been simulated separately \cite{HSU2005177, ANASTASIOU201361, YU201843}, entailing a principle limit in rigor. This work advances the integration of both into a single finite volume simulation by exploiting that the smallest part of the wavelength spectrum does not transmit to the seabed. That is, wave motion not in contact with the seabed does not need to be resolved to depict entrainment.\newline Waves modeled herein do, thus, not encompass the entire eddy length spectrum, which includes motion from all waves, also those that exhibit a minimal perturbation of shear forcing onto the seabed. Therefore, in respect to bottom shear, wave-resolving computation can be attained well before full coastal DNS. Caution is required, as this approach might be valid only for certain wave regimes and climates, and invalid during others that exhibit waves too small to resolve but big enough to be in contact with the seabed.\newline
If conditions are calm with wind waves exhibiting wavelengths much smaller than twice the water depth, then the perturbation of near-bottom tidal currents due to orbital wave motion remains small or insignificant. If conditions are sufficiently agitated such that wavelengths reach in size to the order of the water depth, then wave orbital motion may considerably influence bottom currents. Additionally, high energy waves yield disproportionate increases in erosive fluxes, contributing considerably to shoreline development. Simulating waves and sediment transport in an integrated model, capable of resolving waves, can, thus, enhance accuracy in the depiction of sediment transport.
\newline
The sections below contain the wave-resolving simulation of the Rouse number distribution to resolve fine phenomena in the balance of sediment entertainment and deposition.
The developed Voronoi model is suitable for wave resolution as the variable number of edges per finite volume yields less wave fronts on acute triangle angels. In terms of numerical diffusion \cite{Holleman2013}, Voronoi meshes exhibit a reduction compared to Delaunay meshes \cite{chansolution2018}. An analytical verification of the model, made tractable by dynamic domain contractions, has been documented separately \cite{lawen2023solitary}. Earlier Delaunay versions of the model have been in use since a decade \cite{LAWEN2013330}. 
\newline
Various reasons might have initially supported the choice of unstructured Delaunay triangle meshes \cite{Lawen2014} that followed models based on structured meshes \cite{LAWEN201099} vis-\'{a}-vis Voronoi meshes. For example, cells in the latter mesh type are formed by a varying number of edges. Absent sparse storage mode, arrays for vertices and faces are in size, thus, determined by the Voronoi cell with the highest number of faces. Meanwhile Delaunay cells have always just three horizontal faces, yielding compact arrays for early cache-fitting simulations on multiple bus-snooping and, thus, cache-coherent cores or CPUs.. Voronoi meshing has lately also been applied to oceanography with works \cite{HERZFELD2020101599, RINGLER2013211} mentioning different stability concerns vs. Delaunay meshes, indicating that an algorithm that might be stable on a Delaunay mesh might not necessarily be stable on a Voronoi mesh. \newline The need for the developement of ocean models based on Voronoi meshes has been put forth with dedicated emphasis \cite{RINGLER2013211}: "all 23 global ocean models used in the Intergovernmental Panel on Climate Change (IPCC) 4th Assessment Report \cite{IPCC2007} were based on structured, conforming quadrilateral meshes (see Chapter 8, pg 597 of Randall and Bony, 2007). Our view is that the global ocean modeling community benefits from having a diversity of numerical approaches. While this diversification is well underway with respect to the modeling of the vertical coordinate \cite{HALLBERG199754, BLECK200255}, progress in developing new methods for modeling the horizontal structure of the global ocean on climate-change time scales has lagged behind". The Voronoi mesh-borne model provided herein further contributes to the requested diversity, particularly Voronoi-borne modeling.\newline 
Model validation was conducted with five correlations of simulated surface elevations with time series from a tidal survey of three locations and two seasons. In terms of validated hydrodynamic constituents, reliable validation is frequently relying on time series for surface elevations, as conceded in a number of works \cite{Ming1999, Blumberg1977, oeythree-dimensional1985, parkvertical1993, muintwo-dimensional1996}, and found as the simulated quantity that exhibits the best correlation with survey measurements. This follows from the magnitudes of the hydrostatic- and momentum terms and has been generally observed including in recent works \cite{LAWEN2013330, Lawen2014, YU201757}. That is, earth's gravity nivellates watertables such, that tidal constituents are usually rather well-posed quantities in comparison with velocity components. The bottom drag was calibrated to vertical current profiles to obtain reliable values for bottom roughness.

\section{Method}
\label{sec:unstr}
%For the sake of simplicity, mere momentum advection, that is, the material derivative within the Cauchy momentum description is here arrived at via the conservation of the derived quantity of momentum flux. This also has the advantage to arrive at the governing PDE without the use of a fictitious quantity, that is, stress. 
%https://www.google.com/search?q=is+stress+a+ficticious+quantity&oq=is+stress+a+ficticious+quantity&aqs=chrome..69i57.6302j0j4&sourceid=chrome&ie=UTF-8
%https://math.stackexchange.com/questions/3726461/can-navier-stokes-equation-be-derived-from-cauchy-momentum-equation
The Cauchy PDE adds the depiction of stresses to the Euler momentum PDE. In the Navier-Stokes PDE the stresses of the Cauchy PDE are specified for Newtonian fluids, that is, molecular momentum dissipation being proportional to the fluids shear rate. Reynolds averaged Navier-Stokes (RANS) and large eddy simulations (LES) harness unresolved momentum transport by utilizing the diffusive term in the NS PDE. Approximations and configurations of the latter for coastal ocean domains are known as shallow water PDE or primitive equations, primitive in the sense of fundamental governing function. The model solves, in conjunction with the continuity PDE \ref{pde:continuity}, the incompressible Navier-Stokes PDE \ref{eq:tensor} configured for surface flow, which accounts in the control volume $V$ (\SI{}{\cubic\meter})

\begin{equation}
   \frac{\partial{\left(\rho\:\vec{u}\:V\right)}}{\partial t} +\sum_n\frac{\partial \left(u_i \rho_i\:\vec{u}\:V\right)}{\partial x} = \vec{F} + \vec{\nabla}\cdot\begin{pmatrix}
    \tau_{xx} \tau_{xy} \tau_{xz}\\
    \tau_{yx} \tau_{yy} \tau_{yz}\\
    \tau_{zx} \tau_{zy} \tau_{zz}
    \end{pmatrix}
    \label{eq:tensor}
\end{equation}

for component velocities $\vec{u} = [u\;v\;w]$ (\SI{}{\meter\per\second}), with the force $\vec{F}$ (\SI{}{\kg\meter\per\square\second}) due to the hydroststic pressure gradient and Coriolis acceleration, as explicated in Subsections \ref{subs:material} and \ref{subs:cor}. The stress tensor $\tau_{ij}$ (\SI{}{\kg\per\meter\per\square\second}) is in Subsection \ref{subs:stress} configured for an incompressible Netwonian fluid and horizontally isotropic viscosity. The utilization of finite volume approximations is demonstrated for all terms in the subsequent dedicated Subsections \ref{subs:cont} on continuity, \ref{subs:material} on advection and hydrostatic pressure, \ref{subs:cor} on Coriolis acceleration, \ref{subs:stress} on viscous stress, \ref{subs:smag} on the Smagorinsky coefficient computation, \ref{subs:bound} on hydrodynamic boundary conditions, \ref{subs:sediment} on sediment transport, and \ref{subs:eros} on erosion. These subsections also incorporate a derivation of the respective terms.\newline
If component velocities are uniform throughout a finite volume, then the latter is termed convective: quantities are uniformly "convected" throughout a cell. If component velocities are nonuniform throughout a finite volume, then an algorithm is termed conservative if the same quantities exit and enter adjacent finite volumes through faces. That is, constituents are conserved and not lost throughout the domain.
Variable velocities within one finite volume, that is, the conservative case, correspond to transport velocities remaining in the PDE's derivative. Transport velocities can be arranged outside the PDE's derivative, corresponding to uniform velocity components in a finite volume, if the continuity PDE is inserted into the concerned quantity's PDE. The PDE system is then termed convective. \newline
A uniform velocity is obviously more suitable to warrant constituent emission out of a finite volume. Therefore, conservative algorithms are less likely to ascertain the stability of the simulation, particularly if errors are repetitively amplified in circulations \cite{Lawen2014}. \newline Both cases are here derived via a constituent balance of a finite volume. While considering a convective case of quantity transport, the conservative case is still used to derive the algorithm for the continuity PDE. A control volume balance of quantity flows $j_i$ along $x_i$ ($m$), for dimension $i$ and quantity $f$, yields 
\begin{equation}
    \frac{\partial f  }{\partial t} = -\sum_n{\frac{\partial j_i}{\partial x_i}}dx_i
    \label{eq:trans}
\end{equation}
for $n$ spatial dimensions. That is, for advection with velocity components $u$, $v$, and $w$ ($m~s^{-1}$)
\begin{equation}
    \frac{\partial f  }{\partial t} = -\sum_{[u\;v\;w]}{\frac{\partial \left(u_i f\right)}{\partial x_i}}
\end{equation}

The finite volume approximations utilized to convert the partial differential equations for momentum, continuity, and scalar transport into finite volume equations, are listed in Table \ref{tab:fveapprox} below.

%\begin{table}[htbp]
%\centering
%\caption{\bf Term Approximation}
%\begin{tabular}{ccccccc}
%\hline
%\multicolumn{1}{p{1.3cm}}{\centering Term} &\multicolumn{1}{p{1.3cm}}{\centering Convective\\Upwind}&\multicolumn{1}{p{1.5cm}}{\centering Conservative\\Central\\Difference}&\multicolumn{1}{p{1.1cm}}{\centering Central\\Difference}&\multicolumn{1}{p{1.1cm}}{\centering Diffusive\\Central\\Difference}&\multicolumn{1}{p{1.1cm}}{\centering Total\\Derivative}\\
%\hline
%\multicolumn{1}{p{1.3cm}}{\centering Advection for$\:u,v,w,$\\$h\rho,T,S,c$} & $\nabla$ &  &  &  &  \\
%\multicolumn{1}{p{1.3cm}}{\centering Advection for $q$} & & $\nabla$  & & & \\
%\multicolumn{1}{p{1.3cm}}{\centering Hydro- \\static\\Pressure} & & & $\frac{\partial}{\partial x_i}$ &  &$\frac{\partial}{\partial x_i}$\\
%\multicolumn{1}{p{1.3cm}}{\centering Viscous \\Diffusion} & & & & $\nabla^2$  &  \\
%\multicolumn{1}{p{1.3cm}}{\centering Eddy Shear Rates} & & &$\frac{\partial}{\partial x_i}$& & $\frac{\partial}{\partial x_i}$  \\
%\multicolumn{1}{p{1.3cm}}{\centering $\nabla$\:in\:Eddy \\Viscosity's $\nabla\cdot\tau$}& & &$\frac{\partial}{\partial x_i}$& &$\frac{\partial}{\partial x_i}$  \\
%\multicolumn{1}{p{1.3cm}}{\centering Smagorinsky\\model} & & &$\frac{\partial}{\partial x_i}$& &$\frac{\partial}{\partial x_i}$  \\
%\hline
%\end{tabular}
%\label{tab:fveapprox}
%\end{table}

\newpage
Most of the listed upwind and central difference algorithms have been used before in a similar manner in the 3D Simulation for Marine and Atmospheric Reactive Transport (3D SMART) \cite{LAWEN2013330, Lawen2014} for other transport quantities, such as height $h$ (\SI{}{\meter}), temperature $T$ (\SI{}{\celsius}), salinity $S$ (PSU), and concentration $c$ (\SI{}{\kilogram\per\cubic\meter}), and on triangle meshes instead of Voronoi meshes. 

% Centering the wide table with adjustbox
\begin{table}[ht]
    \centering
    \caption{\bfseries Term Approximation}
    \begin{adjustbox}{center}
        \begin{tabular}{ccccccc}
            \hline
            Term & Convective & Conservative & Central & Diffusive & Total \\[-4pt]
                 & Upwind     & Central      & Difference & Central   & Derivative \\[-4pt]
                 &            & Difference   &            & Difference & \\
            \hline
            Advect $u,v,w,h\rho,T,S,c$ & $\nabla$ &  &  &  &  \\
            Advect $q$ & & $\nabla$  & & & \\
            Hydrostatic Pressure & & & $\frac{\partial}{\partial x_i}$ &  &$\frac{\partial}{\partial x_i}$\\
            Viscous Diffusion & & & & $\nabla^2$  &  \\
            Eddy Shear Rates & & &$\frac{\partial}{\partial x_i}$& & $\frac{\partial}{\partial x_i}$  \\
            $\nabla$ in Eddy Viscosity's $\nabla\cdot\tau$& & &$\frac{\partial}{\partial x_i}$& &$\frac{\partial}{\partial x_i}$  \\
            Smagorinsky Model & & &$\frac{\partial}{\partial x_i}$& &$\frac{\partial}{\partial x_i}$  \\
            \hline
        \end{tabular}
    \end{adjustbox}
    \label{tab:fveapprox}
\end{table}

These approximations of derivatives are expressed in Table \ref{tab:fvealgos} in areas $A$ (\SI{}{\square\meter}) and edges $e$ (\SI{}{\meter}) beside an evaluation of derivatives based on the total differential. $\mathds{1}_{>0}(q)$ denotes the indicator function that evaluated whether the volume flow $q_i$ (\SI{}{\cubic\meter\per\second}) through face $i$ into the finite volume fulfills a particular logical condition such as whether it is bigger than zero, whether the flow is entering or exiting, to facilitate upwinding. $n_i(f)$ denotes here a quantity value at the centroid of a face shared with neighbor $i$ in the neighbor list.\newline
Added to the Wavedyne, the new Voronoi mesh-borne module of the 3D SMART, has been a computation of the derivative based on the total derivative: A polygon's centroid and the centroids of two neighboring cells $\beta$ and $\gamma$ constiute a triangle. That is, the total derivative may be denoted for one of two edges of a triangle between the centroid of a particular cell and the centroids of two adjacent cells, $\beta$ and $\gamma$, if $\beta$ is likewise adjacent to $\gamma$

\begin{equation}
    \delta f_1 = \frac{\partial f_1}{\partial x} \delta x_1 + \frac{\partial f_1}{\partial y} \delta y_1 
\end{equation}
with particular edge  components $\delta x_i$, $\delta y_i$, and $\delta f_i$ for edge $i$ but a common $\partial f/\partial x$ and $\partial f/\partial y$ throughout the triangle, yielding
\begin{equation}
\frac{\partial f}{\partial x} = \left( \delta f_1 - \frac{\partial f}{\partial y} \delta y_1 \right)/ \delta x_1
    \label{eq:dfdx}
\end{equation}
Likewise the total derivative can be denoted for another edge and gradient from Equation \ref{eq:dfdx} inserted
\begin{equation}
    \delta f_2 = \left( \delta f_1 - \frac{\partial f}{\partial y} \delta y_1 \right) \frac{\delta x_2}{\delta x_1} + \frac{\partial f}{\partial y} \delta y_2 
\end{equation}
which may likewise be resolved for the gradient with respect to $y$ 
\begin{equation}
    \frac{\partial f}{\partial y} = \frac{\delta f_2 - \delta f_1 \delta x_2/\delta x_1}{\delta y_2 - \delta y_1 \delta x_2/\delta x_1} 
\end{equation}

In Table \ref{tab:fvealgos} binary $\alpha$ takes a logical functionality, carrying the value $0$ or $1$, and is calculated before the simulation to select edge 1 and 2 such that divisions by small numbers are avoided, enhancing accuracy. Likewise, whereas in Table \ref{tab:fvealgos} the total derivative's $\partial f/\partial y$ is calculated before $\partial f/\partial x$, the calculation is also conducted in inverted order to provide a substitute in case of a division by zero.

\begin{table}[htbp]
\centering
\caption{\bf Finite Volume Approximations}
\begin{tabular}{cc}
\hline
\multicolumn{1}{p{1.5cm}}{\centering } &\multicolumn{1}{p{6cm}}{\centering Integrated Quantity Flux Approximation}\\
\hline
$f'\bigg|_{\substack{\text{upwind}\\{}}}^{\text{convective}}$ & $A^{-1}\sum_i\left[ q_i \left(n_i(f)\mathds{1}_{>0}(q_i) + f_i\mathds{1}_{\leq 0}(q_i) \right)\right]$ \\
$f'\bigg|_{\substack{\text{central}\\{}}}^{\text{conservative}}$ &$\left(2 A\right)^{-1}\sum_i\left[ q_i f_i+n_i(q)n_i(f)\right]$\\
$f'\bigg|_{\substack{\text{central}\\{}}}$ & $\left(2 A\right)^{-1}\sum_i\left[proj_{\perp r}(\vec{e}_i)\left(n_i(f)-f\right)\right]$\\
$f''\bigg|_{\substack{\text{central}\\{}}}$ &  $A^{-1}\sum_i\left[\left|\vec{e}_i\right|\left(n_i(f)-f\right)\right]$\\
$f_y\bigg|_{\substack{\text{total} \\ \text{derivative}\\{}}}$ &  $\frac{\alpha\left(f_{\beta} - f\right) + 0^{\alpha}\left(f_{\gamma} - f \right) -\left(\alpha\left(f_{\gamma} - f\right) + 0^{\alpha}\left(f_{\beta} - f \right)\right)\delta x_2/\delta x_1}{\delta y_2 - \delta y_1 \delta x_2/\delta x_1}$\\
$f_x\bigg|_{\substack{\text{total} \\ \text{derivative}\\{}}}$ & $\left(\alpha\left(f_{\gamma} -f\right)-0^{\alpha}\left(f_{\beta} - f\right) - \delta y_1\partial f/\partial y\right)/\delta x_1$\\
\hline
\end{tabular}
\label{tab:fvealgos}
\end{table}

Arrays for $\alpha$, $\beta$ and $\gamma$ are computed once and for all before the simulation as these arrays depend only on the mesh geometries. Nevertheless, the computational costs of the procedure cannot be assessed as negligible, leading to a doubling of run time vis-\'{a}-vis the central difference approximation listed in Table \ref{tab:fveapprox} and \ref{tab:fvealgos}. Algorithm validation has been conducted with a method of manufactured solutions (MMS), which has been submitted separately for publication \cite{lawen2024navierstokesmanifolds}. The MMS was realized by oscillating the seabed to match the flow field to an analytical solution.
%\cite{https://doi.org/10.48550/arxiv.2104.09183}:
%\begin{equation}
%     \underbrace{\nabla.\left(h\:\mathbf{U }\otimes\mathbf{U }\right)}_{material\:derivative}=\underbrace{\nabla.\left(h\:\nabla.\mathbf{U}\right)}_{eddy\:viscosity} + \underbrace{h\: g \nabla h_E}_{hydrostatic\:pressure} + f_{cor} 
%\end{equation}
%with $\mathbf{U}(\mathbf{r})=[U_1 ... U_{n+1}]^T$, SI units, and the component velocity along time being normalized to one. An individual contracting scale for each spatial dimension, with $U_i \perp H_i$, is not required for obtaining some solutions. Note, in this notation, henceforth, no steady state system is denoted. All dimensions have a velocity, including time, e.g. $U_t, U_x, U_y, U_z$ with $U_t = 1$. In case of incompressible free surface flow the hydrostatic gradient along a vertical coordinate $z$ would be rendered $\partial H_E/\partial z = 0$. 

\subsection{Continuity}
\label{subs:cont}
The continuity PDE is obtained by denoting the transport PDE \ref{eq:trans} for mass. The conservative form is converted into the convective form by applying the product rule for derivatives in the equation for mass continuity.   
\begin{equation}
    \frac{\partial m}{\partial t} = -\sum_n{\frac{\partial \dot{m}}{\partial x_i}dx_i}
\end{equation}
With $\partial m = \partial \left(\rho V\right)$ (\SI{}{\kg\per\cubic\meter} \SI{}{\cubic\meter}) this is rendered to 
\begin{equation}
    \frac{\partial \left(\rho V\right) }{\partial t} = -\sum_n{\frac{\partial \left(u_i\rho V \right)}{\partial x_i}}
\end{equation}
dividing by the polygon area $p$, given $V = ph$, and in a convective form, that is, after application of the product rule
\begin{equation}
    \frac{\partial \left(\rho h\right) }{\partial t} = -\sum_n{u_i \frac{\partial \left(\rho  h \right)}{\partial x_i}}-\rho h\sum_n{\frac{\partial u_i }{\partial x_i}}
\label{pde:continuity}
\end{equation}

In the RHS the first term exhibits the convective form for the quantity $h$ and the second term exhibits the conservative form for a constant quantity $=1$. Henceforth, two corresponding finite volume approximations, for the convective and conservative case respectively, can be inserted from Table \ref{tab:fvealgos}. The first term is approximated with upwinding and the second term with a central difference approximation.\newline $^{+\delta t}$ denotes a quantity at the subsequent time level. Past Delaunay versions of the 3D SMART's species transport \cite{LAWEN2013330, Lawen2014} offered for scalar quantities also semi-implicit matrix reordering algorithms. However, these attained only a tripling of time steps at the expense of Flops for the reordering, rendering the net compute gain questionable.
\[\left(h^{+\delta_t}\rho^{+\delta_t}-h\rho\right)\frac{p}{\delta_t}\]
\begin{equation}
     = \overbrace{\sum_m\left[ \frac{q_i}{h_i} \left(n(h_{i}\rho_{i})\mathds{1}_{>0}(q_i) + h_i\rho_i\mathds{1}_{\leq 0}(q_i) \right)\right]}^{upwind} + \frac{h\rho}{2}\overbrace{\sum_m\left[\frac{ q_i}{h_i}+\frac{n(q_{i})}{n(h_{i})}\right]}^{central \: difference}
\end{equation}

Here $q_i$ is the volume flow through face $i$ based on the component velocities at the cell's centroid. Meanwhile, $n(q_{i})$ is the volume flow based on the component velocities of the neighbor at face $i$. The latter is included to approximate the volume flow at the face between two cells.\newline The total horizontal flows through cell faces parallel to the vertical component velocity $w$ equals the summed up component volume flows. The latter are the products of component velocities and the thereto perpendicular edge components.\newpage That is,  

\begin{equation}
    \frac{q_j}{h_j} = proj_{\perp x}(\vec{e}_j)u - proj_{\perp y}(\vec{e}_j)v
\end{equation}

\subsection{Convective material derivative and hydrostatic pressure}
\label{subs:material}

%\begin{equation}
%    \frac{\partial \left(V q \right) }{\partial t} = \sum_m{j_i(u_i,f_i) q_i}
%\end{equation}
%for $m$ volume flows $j_i$ through faces $f_i = e_i\;h$, edges $e_i$, height $h$, and transience free volume $V = p\;h$ with polygon area $p$ and height $h$.
%\begin{equation}
%    V\frac{\partial q }{\partial t} = \sum_m{ j_i(u_i,f_i) q_i}
%\end{equation}
%Rearranging and dividing by $h$ yields
%\begin{equation}
%    \frac{\partial q }{\partial t} = p^{-1}\sum_m{ j_i(u_i,e_i) q_i}
%\end{equation}
%for nonuniform component velocities inside a finite volume, the conservative case, and
%\begin{equation}
%    \frac{\partial q }{\partial t} = p^{-1}\sum_m{ j_i(u,e_i) q_i}
%\end{equation}
%for uniform component velocities, the convective case. The above PDEs and finite volume equations (FVE) will be utilized to substitute corresponding terms in FVE and PDE. 

The material derivative contains besides the time derivative also advective momentum transport, in the Euler momentum, Cauchy, Navier-Stokes, shallow water, and primitive equations alike. Inserting momentum into the PDE for quantity advection yields

\begin{equation}
    \frac{\partial \left(\vec{u} \rho V \right)}{\partial t} = -\sum_n{\frac{\partial\left(u_i \rho\vec{u}V\right)}{\partial x_i}}
   \label{eq:conservative}
\end{equation}
Applying the product rule to the LHS and RHS yields
\begin{equation}
    \vec{u}\frac{\partial \left( \rho V \right)}{\partial t}+ \rho V\frac{\partial \vec{u}}{\partial t} = -\sum_n\left[u_i \rho V\frac{\partial\vec{u}}{\partial x_i}+\vec{u}\frac{\partial\left(u_i \rho V\right)}{\partial x_i}\right]
\end{equation}
Inserting the conservative form of the continuity PDE into the LHS yields the opportunity to eliminate terms, returning for the LHS and RHS only one term each
\begin{equation}
    \rho V\frac{\partial \vec{u}}{\partial t} = -\rho V\sum_n \left[u_i\frac{\partial\vec{u}}{\partial x_i}\right]
\end{equation}

The fluid from equation \ref{eq:conservative} can be denoted with the velocity in its spatially differential form 
\begin{equation}
   \partial{\left(\rho\:\vec{u}\:V\right)}/\partial t = -\sum_n\frac{\partial \left(\rho_i\:\vec{u}\:V\right)}{\partial x}\frac{dx_i}{dt}
\end{equation}
which matches the form of the RHS of the total derivative $df = \sum\partial f/\partial s\: ds$ with $ds=[dx_1 ...\:dx_n,\:dt]$ divided by the time increment. Hence, 
\begin{equation}
   \frac{d\left(\rho\vec{u}\:V\right)}{dt}=\frac{\partial{\left(\rho\:\vec{u}\:V\right)}}{\partial t} +\sum_n\frac{\partial \left(\rho_i\:\vec{u}\:V\right)}{\partial x}\frac{dx_i}{dt}
\end{equation}
or, in consideration of $u_i = dx_i/dt$
\begin{equation}
 \frac{d\left(\rho\vec{u}\:V\right)}{dt} = \frac{\partial{\left(\rho\:\vec{u}\:V\right)}}{\partial t} + \sum_n\frac{\partial \left(u_i\:\vec{u}\:\rho_i\:V\right)}{\partial x}
\end{equation}
PDE and FVE can, hence, be configured as Euler equation by including forces. As per Newton's second law holds for force $\vec{F}$ and momentum $m\vec{u}$
\begin{equation}
    \vec{F} = d\left(m\vec{u}\right)/d t
\end{equation}
Given $m\vec{u} = \rho\:V\:\vec{u}$, the net force in the LHS is obtained by summing up all forces $F_j$ in a free body diagram. 
\begin{equation}
    \sum{\vec{F_j}} = \sum{\frac{\partial\left( u_i m\vec{u}\right)}{\partial x_i}} + \frac{\partial m\vec{u}}{\partial t}
\label{eq:forcesA}
\end{equation}

In terms of FVE approximation, a quantity balance for a regular or irregularly shaped finite volume returns for quantity $f$ transported with upwind approximation in volume flows $q_i$ into volume $V$
\begin{equation}
    \frac{\partial \left(f V\right) }{\partial t} = \sum_m{\left(q_i \left(n(f_i)\mathds{1}_{>0}(q_i) + f_{i}\mathds{1}_{\leq 0}(q_i) \right)\right)}
    \label{eq:quantity advection}
\end{equation}

$\mathds{1}$ is the indicator function that denoted the logical condition of using quantity values of neighboring cells for faces $i$ where volume flows $q_i$ are positive, i.e. in-flowing. The component velocity basis for the volume flows determines whether this FVE corresponds to the conservative or convective case. The latter is the case if the centroid's component velocities is applied to all faces. Inserting momentum into the FVE for quantity advection \ref{eq:quantity advection} yields

\begin{equation}
    \frac{\partial \left(\vec{u} \rho V \right)}{\partial t} = \sum_i{\left(q_i\left(\rho n(\vec{u}_i)\mathds{1}_{>0}(q_i) + \rho\vec{u}_{i}\mathds{1}_{\leq 0}(q_i) \right)\right)}
\end{equation}

If the force acting on surfaces $i$ of the irregular fluid parcel is pressure, then as per $F_i = (\delta_P )_i proj_{\perp r}(\vec{A}_i)$, with the pressure difference $\delta_P$ and the vector of orthogonal component areas $proj_{\perp r}(\vec{A}_i)$, holds
\begin{equation}
    \sum_i{\left( (\delta_P )_i proj_{\perp r}(\vec{A}_i)\right)} = \sum{\frac{\partial\left( u_i\vec{p}\right)}{\partial x_i}} + \frac{\partial \vec{p}}{\partial t}
\end{equation}

under a regime of a hydrostatic gradient distribution, due to a difference in surface height $\delta_h/2$ between adjacent centroids and the edge of the considered cell, follows

\begin{equation}
    g\sum_i{\left[\left(\rho\frac{\delta_h}{2}\right)_iproj_{\perp r}(\vec{A}_i)\right]} = \sum{\frac{\partial\left( u_i m\vec{u}\right)}{\partial x_i}} + \frac{\partial m\vec{u}}{\partial t}
\end{equation}

\begin{equation}
    g\sum_i{\left[\left(\rho\frac{\delta_h}{2}\right)_iproj_{\perp r}(\vec{A}_i)\right]} = \sum{\frac{\partial \left(u_i\rho V \vec{u}\right)}{\partial x_i}} + \frac{\partial\left( \rho V \vec{u}\right)}{\partial t}
\end{equation}
which is again inserted in the RHS of the FVE. 
\[g\sum_i{\left[\left(\rho\frac{\delta_h}{2}\right)_iproj_{\perp r}(\vec{A}_i)\right]}\]
\begin{equation}
     = \sum_i{\left(q_i\left(\rho n(\vec{u}_i)\mathds{1}_{>0}(q_i) + \rho\vec{u}_{i}\mathds{1}_{\leq 0}(q_i) \right)\right)} + \frac{\partial\left( \rho V \vec{u}\right)}{\partial t}
\end{equation}

\newpage With a discrete time derivative and the term for Coriolis acceleration this yields 
%check sign
\[\frac{\rho V\left( \vec{u}^{+\delta t}-\vec{u}\right)}{\delta t}\]
\begin{equation}
 = 
g\sum_i{\left[\left(\rho\frac{\delta_h}{2}\right)_iproj_{\perp r}(\vec{A}_i)\right]}
 - \sum_i{\left(q_i\left(\rho n(\vec{u}_i)\mathds{1}_{>0}(q_i) + \rho\vec{u}_i\mathds{1}_{\leq 0}(q_i) \right)\right)} + F_c
\end{equation}

The method is first order in space and time to attain high resolution meshes (Figure \ref{fig:wavemesh}) to resolve waves while remaining efficient in terms of Flops: to resolve waves, the cell size should be a log order below the part of the wave spectrum of interest, i.e. maximizing cell count and minimizing Flops per cell.

\subsection{Coriolis acceleration}
\label{subs:cor}
Forces including rotational pseudo forces can be substituted into the left-hand side (LHS) of Equation \ref{eq:forcesA}. For the latter, the LHS has to be transformed into the earth's rotating latitude-longitude reference frame.
To observe the acceleration in a rotating reference frame it can be denoted in terms of the spatial vector relative to the inertial reference frame:

\begin{equation}
    \frac{d^2\vec{r}}{dt}_i = \left[
    \frac{d}{dt} 
    + \vec{\Omega}\times \right]
    \left[\frac{d \vec{r}}{dt} + \vec{\Omega}\times\vec{r} \right]
\end{equation}

where $i$ indicates the inertial reference frame. This yields for all $n$ components denoted in momentum vector $m\vec{u}$:

\begin{equation}
\vec{F}-m \frac{d\vec{\omega}}{dt} \times \vec{r} - 2m \left(\vec{\omega}\times \vec{v}\right) - m \vec{\omega}\times(\vec{\omega}\times \vec{r})
=\frac{\partial m\vec{u}}{\partial t} +\sum_{\vec{r}=[x_1...x_n]}{\left(u_i\frac{\partial m\vec{u}}{\partial x_i}\right)}
\end{equation}

with $\vec{\Omega}$ being earth's angular velocity. This procedure recovers a term to account for earth's rotation alone. The effects due to earth's axial tilt, which is accounted for in the insolation simulation for surface heat exchange, inclination relative to the solar plane, and the sun's inclination relative to the galactic plane, are unanimously deemed negligible at the scale of the required transport accuracy and other uncertainties.
Likewise earth's angular velocity is considered as constant and, hence, its time derivative vanishes:

\begin{equation}
    \vec{F} - 2m \left(\vec{\omega}\times \vec{v}\right) - m \vec{\omega}\times(\vec{\omega}\times \vec{r})=\frac{\partial m\vec{u}}{\partial t} +\sum_{\vec{r}=[x_1...x_n]}{\left(u_i\frac{\partial m\vec{u}}{\partial x_i}\right)}
\end{equation}

The vertical component of the Coriolis acceleration is deemed negligible \cite{kundufluid2016}
and heavily masked by imperfectly defined vertical turbulent momentum transport.
Evaluating the cross products yields for earth's zonal and meridional dimensions a negligible term with quadratic angular velocity, a perpendicular centripetal, radially inward acceleration, absent which 
\begin{equation}
    \vec{F} +    \begin{pmatrix}
    1\\
    -1\\
    0
    \end{pmatrix} 2{}\:m{}\: \omega \sin{\left(\phi\right)} \vec{u} =\frac{\partial m\vec{u}}{\partial t} +\sum_{[x_1 x_2]}{\left(u_i\frac{\partial m\vec{u}}{\partial x_i}\right)}
    \end{equation} 
    
is obtained for a particular latitude $\phi$ ($\SI{}{\radian}$) where earth's angular velocity $\omega$ ($\SI{}{\radian\per\second}$) is given with $2 \pi(24\times60^2$\SI{}{\second}$)^{-1}$. As the Coriolis term does not contain any derivative, no approximation is required.

\subsection{Viscous stress, turbulence, LES, and RANS}
\label{subs:stress}
The consideration of surface forces also accounts for the internal friction of the fluid, the viscosity. Each of the three component velocities, at the centroids of the faces of the examined control volume, undergoes strain in three spatial dimensions, yielding nine strain rate elements which are commonly presented in tensor form as shown in equation \ref{eq:tensor}. Note, the generality of tensor calculus does not need to be exploited in this case and the demonstrated falls into the confines of matrix calculus. The viscous stress tensor is usually denoted in the form below including the Nabla operator from the first-order Taylor expansion to attain the differential notation of the force balance at the infinitesimal control volume. For example, for the first component velocity, the dot product yields for the first tensor row $\partial\tau_{xx}/\partial x + \partial\tau_{xy}/\partial y +\partial\tau_{xz}/\partial z$.
\newline
The Navier-Stokes PDE is set apart from the Cauchy momentum PDE by being specified for Newtonian fluids where -- assuming incompressibility -- stress $\tau_{ij}$ is linearly proportional to the sum of the gradient of velocity $i$ in direction $j$ and the gradient of velocity $j$ in direction $i$. The proportionality coefficient $\mu_{ij}$ is termed the viscosity.
The entirety of all stresses, denoted in the stress tensor, can, hence, be substituted by the proportionality of incompressible Newtonian fluids to the sum of the strain rate tensor and its transpose. Note, the absence of volume viscosity is warranted due to the assumption of incompressibility.
\[\vec{\nabla}\cdot \begin{pmatrix}
    \tau_{xx} & \tau_{xy} & \tau_{xz}\\
    \tau_{yx} & \tau_{yy} & \tau_{yz}\\
    \tau_{zx} & \tau_{zy} & \tau_{zz}
    \end{pmatrix}\]
\begin{equation}
    =\vec{\nabla}\cdot    \left(\begin{pmatrix}
    \mu_{xx}\frac{\partial u}{\partial x} & \mu_{xy}\frac{\partial u}{\partial y} & \mu_{xz}\frac{\partial u}{\partial z}\\
    \mu_{yx}\frac{\partial v}{\partial x} & \mu_{yy}\frac{\partial v}{\partial y} & \mu_{yz}\frac{\partial v}{\partial z}\\
    \mu_{zx}\frac{\partial w}{\partial x} & \mu_{zy}\frac{\partial w}{\partial y} & \mu_{zz}\frac{\partial w}{\partial z}
    \end{pmatrix}
    +\begin{pmatrix}
    \mu_{xx}\frac{\partial u}{\partial x} &  \mu_{yx}\frac{\partial v}{\partial x} & \mu_{zx}\frac{\partial w}{\partial x} \\
    \mu_{xy}\frac{\partial u}{\partial y} & \mu_{yy}\frac{\partial v}{\partial y} & \mu_{zy}\frac{\partial w}{\partial y} \\
    \mu_{xz}\frac{\partial u}{\partial z} & \mu_{yz}\frac{\partial v}{\partial z} & \mu_{zz}\frac{\partial w}{\partial z} \\
\end{pmatrix}
    \right)
    \label{eq:strainrate}
\end{equation}
The RHS contains the strain rate tensor and its transpose. In the case of molecular viscosity, that is, momentum transport due to molecular diffusion and interaction, an isotropic and spatially constant coefficient is assumed for all nine $ij$ combinations. As per the latter assumption, the viscosity coefficient can be denoted outside the tensor.
\begin{equation}
    \vec{\nabla}\cdot   \mu \left(\begin{pmatrix}
    \frac{\partial u}{\partial x} & \frac{\partial u}{\partial y} & \frac{\partial u}{\partial z}\\
    \frac{\partial v}{\partial x} & \frac{\partial v}{\partial y} & \frac{\partial v}{\partial z}\\
    \frac{\partial w}{\partial x} & \frac{\partial w}{\partial y} & \frac{\partial w}{\partial z}
    \end{pmatrix}
    +\begin{pmatrix}
    \frac{\partial u}{\partial x} &  \frac{\partial v}{\partial x} & \frac{\partial w}{\partial x} \\
    \frac{\partial u}{\partial y} & \frac{\partial v}{\partial y} & \frac{\partial w}{\partial y} \\
    \frac{\partial u}{\partial z} & \frac{\partial v}{\partial z} & \frac{\partial w}{\partial z} \\
\end{pmatrix}
    \right)
    \label{term:molecular}
\end{equation}
Inserting term \ref{term:molecular} in equation \ref{eq:strainrate} and resolving its dot product yields for the first row, that is, for the viscous stress term of component velocity $u$
\begin{equation}
\frac{\partial \tau_{xx}}{\partial x}+\frac{\partial \tau_{xy}}{\partial y}+\frac{\partial \tau_{xz}}{\partial z}
=\mu\left(\frac{\partial \left(\frac{\partial u}{\partial x} + \frac{\partial u}{\partial x}\right)}{\partial x}+\frac{\partial \left(\frac{\partial u}{\partial y} + \frac{\partial v}{\partial x}\right)}{\partial y}+\frac{\partial \left(\frac{\partial u}{\partial z} + \frac{\partial w}{\partial x}\right)}{\partial z}\right)
\end{equation}

In Eulerian fluid dynamics, that is, continuum mechanics, the assumption of continuous variables flows directly from the very concept under consideration and is presumed for the quantities' derivatives as well. Therefore, Clairaut's theorem can be applied, rendering the order of partial differentiation immaterial.
This assumption breaks down at quantity jumps. Yet at infinite gradients, Eulerian diffusive models break down anyway. Fortunately, such conditions are, at the considered scale, not present in the described coastal ocean systems.
Therefore, partial derivatives in the LHS can be sorted as follows

\begin{equation}
\frac{\partial \tau_{xx}}{\partial x}+\frac{\partial \tau_{xy}}{\partial y}+\frac{\partial \tau_{xz}}{\partial z}
\mu \left(\frac{\partial^2 u}{\partial x^2}+\frac{\partial^2 u}{\partial y^2} + \frac{\partial^2 u}{\partial z^2}\right)+\mu\frac{\partial \overbrace{\left(\frac{\partial u}{\partial x} + \frac{\partial v}{\partial y}+\frac{\partial w}{\partial z}\right)}^{=0}}{\partial x}
\end{equation}

Here the insertion of the continuity PDE $\nabla\vec{u}=0$ eliminates three of the partial derivatives, yielding for the entire system of PDEs

\begin{equation}
    \frac{\partial{\left(\rho\:\vec{u}\:V\right)}}{\partial t} +\sum_n\frac{\partial \left(u_i\rho_i\:\vec{u}\:V\right)}{\partial x} = 
    \vec{F} +\mu \left(\frac{\partial^2 \vec{u}}{\partial x^2}+\frac{\partial^2 \vec{u}}{\partial y^2} + \frac{\partial^2 \vec{u}}{\partial z^2}\right) 
\end{equation}
Molecular viscosity is isotropic and largely constant and can, therefore, due to being constant, be denoted outside the derivative. Yet, eddy viscosity is not at all constant.
In the case of eddy viscosity, the Smagorinsky model assumes horizontally isotropic viscosity. Furthermore, 
\[\partial w/\partial x<<\partial u/\partial x\] 
\[\partial w/\partial x<<\partial u/\partial y \]
\begin{equation}
\partial w/\partial x<<\partial v/\partial x 
\end{equation}
and the analogous for $\partial w/\partial y$, yielding
\begin{equation}
    \vec{\nabla}\cdot 
    \left(\begin{pmatrix}
    k\frac{\partial u}{\partial x} & k\frac{\partial u}{\partial y} & k_z\frac{\partial u}{\partial z}\\
    k\frac{\partial v}{\partial x} & k\frac{\partial v}{\partial y} & k_z\frac{\partial v}{\partial z}\\
    k\frac{\partial w}{\partial x} & k\frac{\partial w}{\partial y} & k_z\frac{\partial w}{\partial z}
    \end{pmatrix}
    +\begin{pmatrix}
    k\frac{\partial u}{\partial x} &  k\frac{\partial v}{\partial x} & 0\frac{\partial w}{\partial x} \\
    k\frac{\partial u}{\partial y} & k\frac{\partial v}{\partial y} & 0\frac{\partial w}{\partial y} \\
    k_z\frac{\partial u}{\partial z} & k_z\frac{\partial v}{\partial z} & k_z\frac{\partial w}{\partial z} \\
\end{pmatrix}
    \right)
\end{equation}
with horizontally isotropic eddy viscosity $k$ and vertical eddy viscosity $k_z$. 
Therefore, Clairaut's theorem cannot be applied except for the third row of the transpose. That is, the transpose of the strain rate tensor remains relevant. The partial derivatives are, thus, collected differently, yielding for the horizontal component velocities
\[\frac{\partial{\left(\rho\:u\:V\right)}}{\partial t} +\sum_n\frac{\partial \left(u_i\rho_i\:u\:V\right)}{\partial x}\]
\begin{equation}
 = 
    \vec{F} + 2\frac{\partial \left(k\partial u/\partial x\right)}{\partial x}+\frac{\partial \left(k\left(\partial u/\partial y+\partial v/\partial x\right)\right)}{\partial y}+\frac{\partial \left(k_z\partial u/\partial z\right)}{\partial z}
\end{equation}

\[\frac{\partial{\left(\rho\:v\:V\right)}}{\partial t} +\sum_n\frac{\partial \left(u_i\rho_i\:v\:V\right)}{\partial x}\]
\begin{equation}
 = 
    \vec{F} + 2\frac{\partial \left(k\partial v/\partial y\right)}{\partial y}+\frac{\partial \left(k\left(\partial v/\partial x+\partial u/\partial y\right)\right)}{\partial x}+\frac{\partial \left(k_z\partial v/\partial z\right)}{\partial z}
\end{equation}
The dot product of the transpose of the strain rate tensor for the third row, the vertical component velocity $w$:
\begin{equation}
    \vec{\nabla}\cdot\left(k\frac{\partial w}{\partial x}+k\frac{\partial w}{\partial y}+k_z\frac{\partial w}{\partial z}\right)+\frac{\partial \left(k_z\frac{\partial u}{\partial z}\right)}{\partial x}+\frac{\partial \left(k_z\frac{\partial v}{\partial z}\right)}{\partial y}+\frac{\partial \left(k_z\frac{\partial w}{\partial z}\right)}{\partial z}
\end{equation}
Applying the product rule and sorting terms produces
\[\vec{\nabla}\cdot\left(k\frac{\partial w}{\partial x}+k\frac{\partial w}{\partial y}+k_z\frac{\partial w}{\partial z}\right)\]
\begin{equation}
    +\frac{\partial k_z}{\partial x}\frac{\partial u}{\partial z}+\frac{\partial k_z}{\partial y}\frac{\partial v}{\partial z}+\frac{\partial k_z}{\partial z}\frac{\partial w}{\partial z}+k_z\left(\frac{\partial \frac{\partial u}{\partial z}}{\partial x}+\frac{\partial \frac{\partial v}{\partial z}}{\partial y}+\frac{\partial \frac{\partial w}{\partial z}}{\partial z}\right)
\end{equation}
For the last three terms, Clairaut's theorem can again be applied, changing the order of partial differentiation. Furthermore, some terms are approximated with finite differences.
\[    \vec{\nabla}\cdot\left(k\frac{\partial w}{\partial x}+k\frac{\partial w}{\partial y}+k_z\frac{\partial w}{\partial z}\right)\]
\begin{equation}
+\frac{\delta k_z}{\delta x}\frac{\delta u}{\delta z}+\frac{\delta k_z}{\delta y}\frac{\delta v}{\delta z}+\frac{\delta k_z}{\delta z}\frac{\delta w}{\delta z}+k_z\frac{\partial\overbrace{\left(\frac{\partial u}{\partial x}+\frac{\partial v}{\partial y}+\frac{\partial w}{\partial z}\right)}^{=0}}{\partial z}
\end{equation}
yielding again the opportunity to exploit the continuity PDE to eliminate terms. Also, the finite difference approximation yields the opportunity to rearrange divisors
\begin{equation}
\vec{\nabla}\cdot\left(k\frac{\partial w}{\partial x}+k\frac{\partial w}{\partial y}+k_z\frac{\partial w}{\partial z}\right)
+\frac{\delta k_z}{\delta z}\overbrace{\left(\frac{\delta u}{\delta x}+\frac{\delta v}{\delta y}+\frac{\delta w}{\delta z}\right)}^{\approx 0}
\end{equation}
recovering the finite difference approximation of the continuity PDE which is approximately zero and, thus, eliminating further terms.
Inserted into the PDE for the vertical component velocity this yields
\[    \frac{\partial{\left(\rho\:w\:V\right)}}{\partial t} +\sum_n\frac{\partial \left(u_i\rho_i\:w\:V\right)}{\partial x}\]
\begin{equation}
 = \vec{F} +\frac{\partial \left(k\partial w/\partial x\right)}{\partial x} +\frac{\partial \left(k\partial w/\partial y\right)}{\partial y}+ \frac{\partial \left(k_z\partial w/\partial z\right)}{\partial z}
   \label{eq:NSpde}
\end{equation}
\newline
Table \ref{tab:fveapprox} and \ref{tab:fvealgos} list the selectable approximations for the eddy diffusive terms.
Transport unresolved by the mesh is at the smallest scale molecular diffusion that is present also during laminar flow. Negligible molecular diffusion but also unresolved eddies are usually modeled as random, that is, diffusive phenomena and referred to by the fictitious quantity eddy diffusion.\newline Direct numerical simulation (DNS) is computationally prohibitively expensive for geophysical models absent a quantum computing resource \cite{itani-hal-03129398, bharadwajintroduction2022}. 
Henceforth, unresolved turbulence is treated with two baroque chains of fictitious models and quantities: 1.) Reynolds averaged Navier Stokes (RANS) and 2.) large eddy simulations (LES).
In the first approach, transient fluctuations are split off the modeled velocity, whereas, in the second, transport is filtered by deducting   \cite{1963MWRv...91...99S} spatially unresolved transport.
I.e. the former smooths over time, the latter smooths over space, and hybrid models have been developed as well. 
\newline
The following subsections show that the RANS approach is more receptive to account for buoyancy which is important for vertical turbulence whereas LES, developed by Smagorinsky, is computationally more effective for isotropic flows which are a valid description of the horizontal. Therefore, commonly RANS models, in particular k-$\epsilon$ models, are used for the vertical and an LES-Smagorinsky model for the horizontal. The same approach has been selected for this model and documented in subsequent sections.\newline The transport PDE derived above, including a diffusive term for turbulence and in conjunction with a density quantifying PDE is termed the primitive equations where primitive corresponds to its basic governing utility and not a lack in sophistry. Or, if inertial considerations are restricted to two dimensions and the vertical is approximated with mere continuity, then the yielded set of PDE is termed shallow water equations. 

\subsection{Smagorinsky turbulence}
\label{subs:smag}
Transport of vector and scalar quantities is deemed split into resolved and an unresolved components. This step is usually referred to as filtering with a filtered velocity $\bar{u}$ and an unresolved residual velocity. %The deviatoric stress $\tau_{ij} = \rho \bar{u_i^* u_j^*}$ for the fluctuating velocity $\mathbf{u^*}$. 
Underlying the Smagorinsky model is the assumption that Reynolds stress can be modeled with the rate of strain tensor, introducing another ficticious viscosity that is itself modeled by the rate of strain tensor. The Smagorinsky model can has inherent the assumption of an even weighting of the fluctuating velocity. The averaging in the filter function can also introduce a heavier weight close to the centroids given that the velocities are stored at the centroid. 
\newline
The Smagornisky model corresponds only to a uniform isotropic filter that is also termed box filter. Gaussian bell shaped filter function exists as well that particularly sample the fluctuation velocity at the centroids while still smoothing fluctuations. 
The invented eddy viscosity is dependent on the filter size, here the Voronoi cell size. Molecular viscosity is usually ignored given that it is logarithmic orders of magnitudes smaller than eddy viscosity.
The Smagorinsky model is simpler than the RANS model by assuming an isotropic turbulence which is warranted in the horizontal plane but not for the vertical plane.\newline In the Smagorinsky model, the strain rate tensor of the resolved flow represents the local deformation of the flow. The model is obtained from Kolmorov modeling \cite{kolmogorov1941local} of the turbulent energy dissipation from the average of the fluctuating velocity where a proportionality of the former to the Reynolds stress is inferred from dimensional analysis. The utilization of a diffusive term under the assumption of random and, thus, gradient driven transport, is readily obtained from the fluid's Lagrangian description given that random transport holds in the thermodynamic Lagrangian domain at every scale of unresolved geometry.
\begin{equation}
    k_H =\frac{C_S P}{2} \sqrt{\left(\frac{\partial u}{\partial x}\right)^2+\left(\frac{\partial v}{\partial y}\right)^2+2^{-1}\left(\frac{\partial v}{\partial x}+\frac{\partial u}{\partial y}\right)^2}
    \label{eq:cs}
\end{equation}
with the Smagorinsky coefficient $C_S$. The eddy diffusivity constant is usually \cite{yangmodeling2008, chenunstructured-grid2011} horizontally treated as isotropic but computed separately for the vertical.  The eddy diffusivity is for the momentum transport the eddy viscosity $\mu$ which can be set for the horizontal or vertical as prognostic or diagnostic. That is, in the latter case a value obtained from a parameter identification is applied. Otherwise, in the prognostic case the horizontal eddy viscosity is computed with the Smagorinsky model.\newline
Numerical diffusion is to be expected as well, that is, artificial diffusion inherent to continuum descriptions of physics. Numerical diffusion does not threaten the stability but accuracy of a simulation. Nevertheless, LES and RANS models permit nominal compliance with the Navier-Stokes PDE, comprehensiveness in approximating all terms, or may entail stability benefits that a diffusive term conveys to some algorithms. \newline
The Smagorinsky model has been defined as an L2-norm of derivatives of the component velocities. The evaluation of the Cartesian gradient at the centroids is not readily available in unstructured meshes and requires dedicated computation \cite{AMultiscaleNonhydrostaticAtmosphericModelUsingCentroidalVoronoiTesselationsandCGridStaggering}. 
Given that magnitudes of other uncertainties are of higher logarithmic orders in LES and RANS models, it appears questionable to expend Flops to refine the computation of the derivatives in Eq. \ref{eq:cs} to compute estimates for this fictitious quantity.
Nevertheless, to not have to quantify the uncertainty due to these interpolations, the derivatives of the velocity components at the centroid are here rigorously computed by applying the total derivative to each triangle spanned between a centroid and two adjacent neighbors. 
The most of this computation can be computed ahead of the simulation and stored in auxiliary arrays as provided in Table \ref{tab:fvealgos}. 
\newline
The filtered transport in the eddy diffusive terms requires a proportionality coefficient to link the strain rate to an eddy diffusive force. In coastal ocean momentum transport, it is commonly assumed that this coefficient grows with the LES spatial filter, represented by the Voronoi polygon size $P$, and the strain rate terms in the root of Equation \ref{eq:cs} above.
Algorithms to compute gradients are listed in Table \ref{tab:fveapprox} and \ref{tab:fvealgos}. All geometric meta-quantities are calculated in advance ahead of the simulation and stored to minimize the computational load.

\subsection{Hydrodynamic boundary conditions}
\label{subs:bound}
For vector quantities, that is, velocity components, the hydrostatic pressure gradient ascertains that each finite volume with a no-flux face does not grow out of bounds. The no-flux boundary is simply attained by setting all fluxes through the boundary face to zero. For scalar quantities, that is, quantities absent a hydrostatic pressure term, such as temperature, salinity, sediment, and tracer, the free-slip boundary condition ascertains that quantities do not grow out of bounds. The free-slip boundary condition ascertains that the velocity vector in the finite volume is parallel to the boundary, that is, there is no component perpendicular to a solid boundary face. \newline
At open boundary conditions that connect the model to the open sea, the surface elevation is set to the surveyed tidal meter time series or tidal data from any other source.
%file:///C:/Users/Edwin/Downloads/Bottom_friction_models_for_shallow_water_equations.pdf) 
\newline
Rigor in depicting bottom stress is limited by usually unknown roughness distribution of the vast seafloor. Some assumptions have heretofore though been made: Absent a velocity to shed from or fluid to interact with, no friction force can act, rendering plausible the assumption of bottom stress being dependent on velocity and density. If an advected fluid particle encounters and is exerted upon by a roughness element, then the number of encounters is, absent orientation of the roughness element, in such Lagrangian model proportional to the advecting velocity's magnitude and particle density. Furthermore, the net force exerted can, as any force, be denoted in its components, that is, in terms of velocity components as momentum is removed from the same. Therefore, the bottom stress boundary proportionality is obtained and rendered an equation \cite{feddersendrag2003, fariavertical1998} by introducing a proportionality coefficient  
\begin{equation} 
\vec{\tau} \propto \rho\langle \|\vec{u}\|\vec{u}\rangle
\end{equation}

\begin{equation} 
\vec{\tau} = \rho c_d \langle \|\vec{u}\|\vec{u}\rangle
\label{eq:tau}
\end{equation}
with the coefficient $c_d$ being termed drag coefficient.
Evident with Equation \ref{eq:tau} becomes, that for a particular stress $\vec{\tau}$, the drag coefficient $c_d$ depends on the height above the bottom given the velocity's dependency on the same. For example, if a slim bottom layer is modeled, utilizing a slow segment within the vertical velocity profile, then the drag coefficient grows.
\newline
An estimate for $c_d$ is usually obtained \cite{grantcombined1979} by applying the Prandtl-K\'arm\'an logarithmic velocity profile to oceanic applications, that is,

\begin{equation}
    \|\vec{u}\| = \frac{u_{*}}{\kappa}\ln{\left(\frac{z}{z_0}\right)}
\label{eq:logvel}
\end{equation}

where the friction velocity is defined with
\begin{equation}
    \|\vec{\tau}\| = \rho u_{*}^2
    \label{eq:friction}
\end{equation}
with the bottom stress $\vec{\tau}$, the friction velocity $u_{*}$, the K\'arm\'an constant $\kappa = 0.4$, the roughness length $z_0$, and the height above the seafloor $z$. Bottom drag has been estimated and modeled by conducting parameter identification for the friction velocity and drag coefficient \cite{fariavertical1998}, or by identifying a suitable roughness length \cite{isobeestimate2006, weisbergcirculation2006, yangmodeling2008}. Yet, whereas the drag coefficient depends on the bottom layer height, the friction velocity dependents on the local velocity. Both are, thus, not independent of transient quantities. Therefore, for the purpose of modeling various depths and flow regimes, only the roughness length is sufficiently fundamental to hold for the entire tidal cycle. Resolving Equation \ref{eq:logvel} for $u_{*}$ and specifying the RHS for two heights, eliminates the friction velocity

\begin{equation}
    \frac{\kappa\|\vec{u_1}\|}{\ln{\left(\frac{z_1}{z_0}\right)}} =     \frac{\kappa\|\vec{u_2}\|}{\ln{\left(\frac{z_2}{z_0}\right)}}
\end{equation}

and, after rearrangements, resolves to the ubiquitous form for roughness length identification in atmospheric and oceanic applications alike: 

\begin{equation}
\ln{z_0} =         \frac{{\|\vec{u_1}\|\ln{z_2}} -    \|\vec{u_2}\|\ln{z_1}}{\|\vec{u_1}\| -     \|\vec{u_2}\|}
\label{eq:zid}
\end{equation}

The drag coefficient $c_d$ in the boundary condition is then obtained by taking the magnitude of the LHS- and RHS-vector in Equation \ref{eq:tau} whilst substituting for the LHS the RHS of Equation \ref{eq:friction}.

\begin{equation} \rho u_{*}^2 = \|\rho c_d \langle \|\vec{u}\|\vec{u}\rangle\|
\label{eq:magofvec}
\end{equation}

Furthermore, Equation \ref{eq:logvel}, resolved for the friction velocity, substitutes the same in Equation \ref{eq:magofvec}. 

\begin{equation} \left(\frac{\kappa \|\vec{u}\|}{ln{{z}-ln{z_0}}}\right)^2 =  c_d  \|\vec{u}\|\|\vec{u}\|
\end{equation}
Thus, yielding the common form \cite{chenunstructured-grid2011} for calculating the drag coefficient distribution from the roughness length
\begin{equation} 
c_d(x,y,t) = \left(\frac{\kappa }{ln{{z(x,y,t)}-ln{z_0}}}\right)^2  
\label{eq:cd}
\end{equation}

Denoted in Equation \ref{eq:cd} is explicitly, that $c_d(x,y,t)$ is not a constant but a horizontal distribution, if the layer thickness $z(x,y,t)$ is not constant. Seabed change or tidal transience, depending on the layer configuration, can result in a time dependency. 

\subsection{Sediment transport and settling}
\label{subs:sediment}
Sediment is transported like any other constituent with the addition of sediment settling and bottom sediment entrainment.
A mass balance for the sediment type C returns

\[    \frac{\partial C}{\partial t} + \frac{\partial (u C)}{\partial x} + \frac{\partial (v C)}{\partial y} + \frac{\partial \left((w-w_s) C\right)}{\partial z} \]
\begin{equation}
    = \frac{\partial}{\partial x}\left(D_x \frac{\partial C}{\partial x}\right) + \frac{\partial}{\partial y}\left(D_y \frac{\partial C}{\partial y}\right) + \frac{\partial}{\partial z}\left(D_z \frac{\partial C}{\partial z}\right) - E
\end{equation}

with the entrainment term E.
The consideration for eddy diffusion in the chapters above have shown that the horizontal eddy diffusive coefficients are horizontally isotropic. Furthermore, the insertion of continuity permits to denote the velocities outside of the derivatives. 

\[    \frac{\partial C}{\partial t} + u\frac{\partial C}{\partial x} + v\frac{\partial C}{\partial y} + \left(w-w_s \right)\frac{\partial C}{\partial z} \]
\begin{equation}
    = \frac{\partial}{\partial x}\left(A_h \frac{\partial C}{\partial x}\right) + \frac{\partial}{\partial y}\left(A_h \frac{\partial C}{\partial y}\right) + \frac{\partial}{\partial z}\left(D_z \frac{\partial C}{\partial z}\right) - E
\end{equation}

The settling velocity $w_s$ for fine sediment, set in the input file, can be calculated with  Stoke's law 
\begin{equation}
    w_s = \frac{d^2  g (\rho_s - \rho)}{18 \mu}
    \label{eq:settling}
\end{equation}
    
where $d$ is the diameter of the sediment particle (\SI{}{\meter}), $g$ is the acceleration due to gravity (\SI{}{\meter\per\square\second}), $\rho_s$ is the density of the sediment particle (\SI{}{\kilogram\per\cubic\meter}), $\rho$ is the density of the fluid (\SI{}{\kilogram\per\cubic\meter}), and $\mu$ is the dynamic viscosity of the fluid (\SI{}{\kilogram\per\meter\per\second}). 
\newline
Obviously, no sediment passes through the surface or horizontal land boundaries. The governing boundary condition is, hence, the bottom boundary condition as detailed in Section \ref{sec:waves} on waves and erosion.

\subsection{Waves and entrainment}
\label{subs:eros}
The influence of orbital wave motion on near-bottom tidal currents depends on the length of agitating waves. If conditions are calm with wind waves exhibiting wavelengths much smaller than twice the water depth, then the perturbation of near-bottom tidal currents due to orbital wave motion remains small or insignificant. If conditions are sufficiently agitated such that wavelengths of high energy reach in size to the order of the water depth, then wave orbital motion considerably governs bottom currents. Additionally, high energy waves yield disproportionate increases in erosive fluxes, contributing considerably to shoreline development. The Rouse number that indicates whether sediment entrainment or deposition occurs, is given with

\begin{equation}
    Ro = \frac{w_s}{\kappa u_*}
\end{equation}

The survey current meter recorded for sheltered conditions wavelengths not exceeding $0.5$ (\SI{}{\meter}). Therefore, erosive fluxes have been simulated for three conditions with a dedicated high resolution mesh for wave-resolving simulations as detailed in Subsection \ref{sec:waves} of Section \ref{sec:application}: tidal currents during sheltered conditions and during two high energy wave scenarios. Waves are approximated as second order Stokes waves as per the Le M\'ehaut\'e diagram.

%The survey consistently found TSS levels on the order of $10$ $mg/l$. The seabed change distribution is the difference between diagnostic settling based on surveyed TSS and settling velocity $w_s$ from Equation \ref{eq:settling} and the distribution of erosive fluxes based on simulations conducted with the wave resolving mesh.

\section{Application}
\label{sec:application}

\begin{figure}[h]%[htbp]
    \centerline{
    }
    \includegraphics[width=1.0\linewidth]{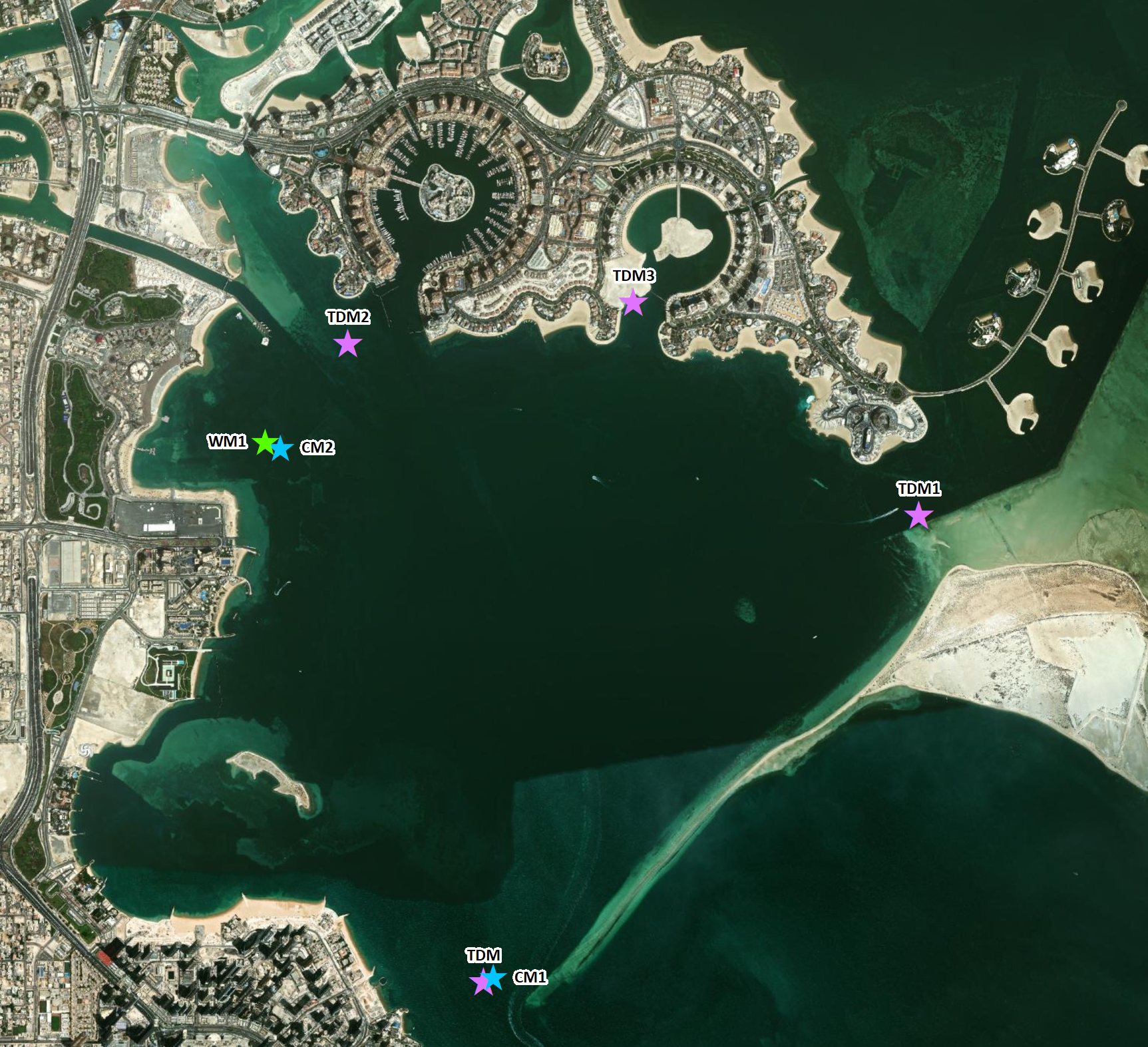}
    \caption{Doha Bay.}
\label{fig:bay}
\end{figure}

Documentation to step-by-step build the model is given in below subsections on meshing \ref{sec:meshing} and case dependent horizontal as well as vertical boundary conditions \ref{sec:bc}. That is, setting tidal boundary conditions  to drive the model and the bottom friction parameter to attenuate it respectively. %Commands and specifications entered for simulated scenarios are also provided as an \hyperref[sec:code]{appendix}. 
The water body modeled is shown in Figure \ref{fig:bay}. A highly resolved beach model for wave simulations is nested within a model for the entirety of the bay. 

Five survey locations in Doha Bay provided for boundary forcing (two) and triple validation, besides examining the same for two seasons.

\subsection{Meshing}\label{sec:meshing}
%\begin{figure*}[b] %b for bottom
%\includegraphics[width=1.0\linewidth]{AsIsMesh.png};
%\caption{Mesh for the study site before the groyne modification.}
%\label{fig:AsIsMesh}
%\end{figure*}

%\begin{figure*}[b] %b for bottom
%\centering
%\includegraphics[width=1.0\linewidth]{PostBathyWithUneditedMarina.png};
%\caption{Bathymetry for state after the development with unedited Marina bathymetry.}
%\label{fig:PostBathy}
%\end{figure*}

\begin{figure}[h] %b for bottom
\centering
\begin{adjustbox}{center}
\pgfplotsset{compat=1.17, width=1.0*\linewidth}
\begin{tikzpicture}
    \begin{axis}[
    xlabel={Eastern UTM/[\si{\meter}]},
    ylabel={Northern UTM/[\si{\meter}]},
        xmin=551317.8,
        xmax=558837.2, 
        ymin=2801017.9,
        ymax=2806951,
        xtick={552000, 553000, 554000, 555000, 556000, 557000, 558000}, 
        xticklabels={ $553{,}000$, $553{,}500$, $554{,}000$, $554{,}500$, $555{,}000$, $555{,}000$, $555{,}000$},
        ytick={2801500, 2802000, 2802500, 2803000, 2803500, 2804000, 2804500, 2805000, 2805500, 2806000, 2806500}, 
        yticklabels={$2{,}801{,}500$, $2{,}802{,}000$, $2{,}802{,}500$, $2{,}803{,}000$,
        $2{,}803{,}500$,
        $2{,}804{,}000$, $2{,}804{,}500$, $2{,}805{,}000$, $2{,}805{,}500$,
        $2{,}806{,}000$,
        $2{,}806{,}500$},
        tick style={draw=none}, 
        axis line style={draw=none},
        width=1*\linewidth,
        height=.7874*\linewidth,
        scaled ticks=false,
        clip = false,
    ]   

    \addplot graphics [                   xmin=551317.8,
        xmax=558837.2,
        ymin=2801017.9,
        ymax=2806961] {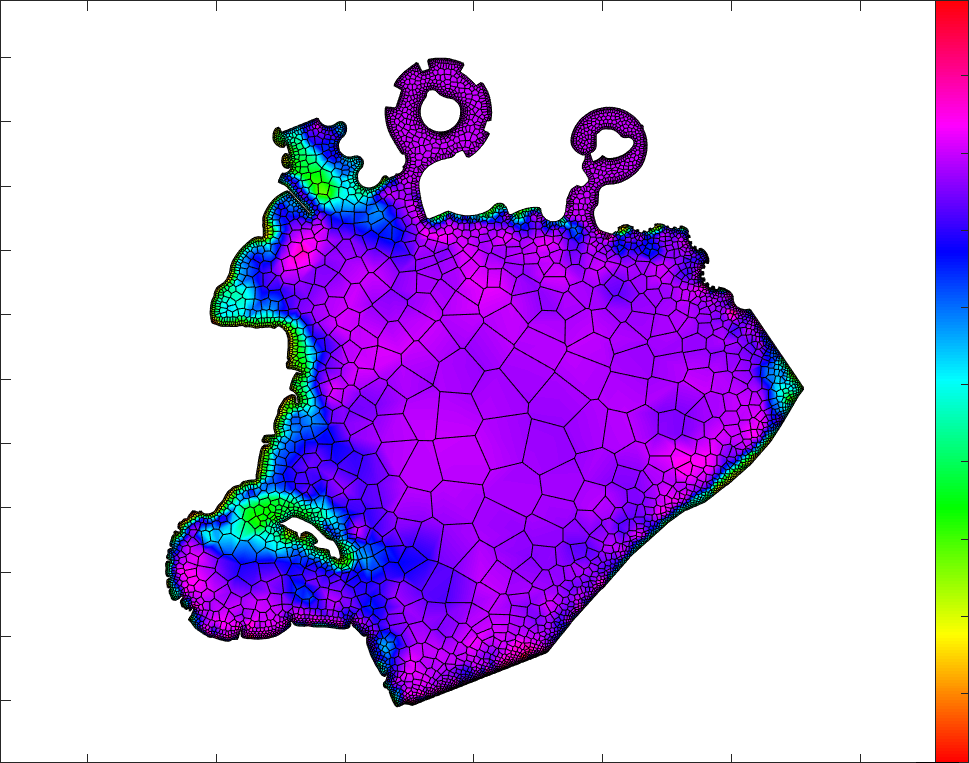};
        
    \end{axis}  
    \begin{axis}[ylabel={Bathymetry/[\si{\meter}]}, 
    hide x axis,
    major tick length=0pt,
    minor tick length=0pt, 
    xmin=0.0, xmax=5, 
    ytick={.5, 1, 1.5, 2, 2.5, 3, 3.5, 4, 4.5},
    yticklabels={$.5$, $1$, $1.5$, $2$, $2.5$, $3$, $3.5$, $4$, $4.5$},
    tick style={draw=none}, 
    axis line style={draw=none},
    ymin=0.05, ymax=4.97, clip=false, axis y line*=right,  ylabel near ticks,
    width=1*\linewidth,
    height=.7874*\linewidth
    ]
    \end{axis}
    \end{tikzpicture}
    \end{adjustbox}
    \caption{Bathymetry with corrected Marina design depth. The wave-resolving mesh is shown in Figure \ref{fig:wavemesh}.}
\label{fig:AsIsBath}
\end{figure}

To obtain the horizontal geometry of the sea surface, a satellite image, as shown in Figure \ref{fig:bay}, can be downloaded from Google Earth or any other image, map, or CAD drawing be used.
The coastline and boundary are then marked with a 24-bit RGB code identifier in a .bmp file in any .bmp editing tool. All land pixels are then automatically flood-filled after setting all other pixels to $0$ with the logical array and flood fill function  \verb+excise('image.bmp',RGB)+ with a particular 24-bit \verb+RGB+ color, that is, with maximal component values \verb+[255 255 255]+. 
\newline
Maps and CAD designs of future developments can be superimposed with the script  \verb+overlay+. 
The mesh is created directly from the .bmp with the mesh generator  \verb+meshing22a('image.bmp')+. The latter automatically provides for a higher resolution at the boundary between land and sea by first distributing Voronoi polygon seeds by sweeping along the shore with increasing distances followed by three iterations of mesh relaxation. The relaxation algorithm redistributes as per Lloyd's algorithm but is based on a discrete tesselation.
The mesh, depicted in Figure \ref{fig:AsIsBath}, is geo-referenced by marking two reference coordinates, ($x_a, y_a$) and ($x_b, y_b$) within the 24-bit .bmp file fed to the meshing generator with a particular 24-bit \verb+RGB+ color.\newline
The function  \verb+[x1,y1,xd,yd] = coord23('.bmp',RGB,xa,ya,xb,yb)+ then returns the coordinates of the bottom left corner (\verb+x1,y1+) of the image and the first reference point (\verb+xa,ya+) in pixel coordinates (\verb+xd,yd+). These can be used to geo-reference mesh coordinates $\mathbf{r} =[x\; y]$ for local sites if spherical coordinates are not entertained. That is, $\mathbf{r}_i = \mathbf{r}_{1}^{UTM}+\left(\mathbf{r}_{a}^{UTM}-\mathbf{r}_{1}^{UTM}\right)\mathbf{r}_i^{bmp}/\mathbf{r}_{a}^{bmp}$ with the coordinate of the bottom left corner $\mathbf{r}_1$ and the reference point $\mathbf{r}_a$.
\newline
The latter georeferencing is embedded within the script for the bathymetry interpolation onto the mesh \verb+bath22a+ by bookkeeping for each cell a depth value in the vector \verb+hb+. The interpolation of \verb+bath22a+ attributes surveyed and remotely sensed depths as per their area share of Voronoi polygons. The composite of survey and remotely sensed bathymetry is shown in Figure \ref{fig:AsIsBath}.
\newline
A string of finite volume cells that bound mangrove nooks, ongoing developments, coverage by marine vessels, or ill-resolved harbor bathymetry can be identified in an index-plot with \verb+plot0(u,v,1:length(u),'mesh.mat','0','-','txt')+, null vectors as dummy velocities, and the visualized quantity, here the indices, being denoted (\verb+'txt'+) in each delineated (\verb+'-'+) cell.
The meshing code is not discussed here because the consistency of the output Voronoi diagram can be ascertained visually.

The identified boundary of the area to be amended is then denoted in a list of cells (\verb+list_const+) with a constant null concentration (\verb+c_const = 0+) in a simulation's case file (\verb+c(list_const) = c_const+). Any random cell inside a listed area of concern (\verb+list_c0+) can conveniently return via advective propagation all the area's cells as non-zero (\verb+run('case_file.m')+).

That is, the tidal advection simulation is merely exploited to automatically mark the area
delineated by \verb+list_const+. Identified cells, that is, a concerned area \verb+ area_i = (c>0)+, may be book-kept \verb+save name.mat area_i+ and set (\verb+hb(area_i) = +) to the desired depth, in this case, $4 m$ for the harbor and $1 m$ for the mangrove forest.

\subsection{Boundary conditions}
\label{sec:bc}
\begin{figure}[h]
    \centering
    \begin{minipage}{0.8\linewidth}
        \begin{tikzpicture}
            % First axes: Plot the image and rectangle, but make the axes invisible
            \begin{axis}[
    xlabel={Set Reference Distance $z_1$ from Bottom/[\si{\meter}]},
    ylabel={Identified Roughness Length/[\si{\meter}]},
    xmin=1.1,
    xmax=1.4,
    ymin=0.16,
    ymax=0.34,
    xtick={1.15, 1.2, 1.25, 1.3, 1.35, 1.4},
    xticklabels={$1.15$, $1.2$, $1.25$, $1.3$, $1.35$, $1.4$},
    ytick={0.16, 0.18, 0.2, 0.22, 0.24, 0.26, 0.28, 0.3, 0.32, 0.34},
    yticklabels={$0.16$, $0.18$, $0.2$, $0.22$, $0.24$, $0.26$, $0.28$, $0.3$, $0.32$, $0.34$},
    tick style={draw=none}, 
    axis line style={draw=none},
    width=\linewidth,
    height=0.8\linewidth,
    axis equal image % Ensure image aspect ratio fits
]
            % Add the main image
            \addplot graphics [xmin=1.1, xmax=1.4, ymin=0.16, ymax=0.34] {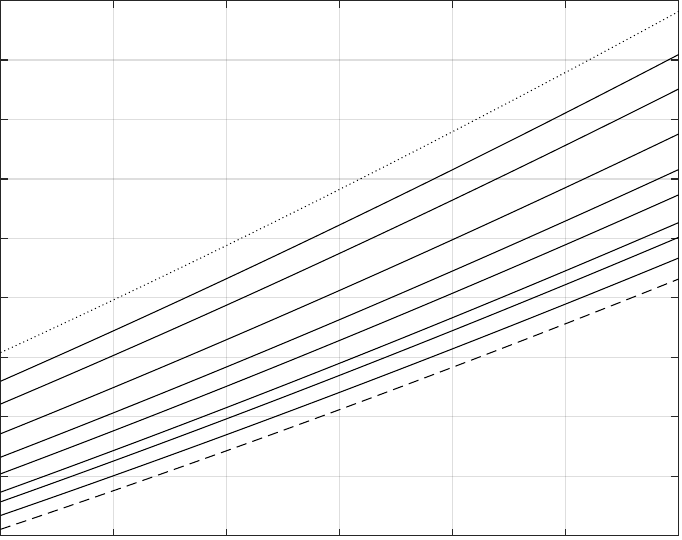};
         
            \end{axis}
        \end{tikzpicture}
    \end{minipage}
    \caption{Identified roughness length $z_0$ vs. varied uncertain reference height $z_1 = 1.25 \pm$ \SI{0.15}{\meter} and altered minimal velocity difference $\delta u$ from \SI{1}{\percent} (dashed) to \SI{10}{\percent} (dotted) with ascending one-percent increments in between (solid).}
    \label{fig:ZID}
\end{figure}

Tidal, current, open, and/or other horizontal boundaries are specified by plotting cell indices, with the accordingly configured plotting function, with a zero-order interpolation \verb+plot23(u,v,1:length(u),'mesh.mat','0','-','txt')+. The mesh generator numbers cell indices sequentially along the boundary. Therefore, quantities at the boundary can be set to boundary conditions by referring to a particular boundary section with \verb+index_1:index_2+. For example, in the case of a tidal boundary condition \verb+nodes(1:length(istart:iend)) = istart:iend+. 
\newline
The bed roughness length was identified, as illustrated in Figure \ref{fig:ZID}, with Equation \ref{eq:zid} to specify the bottom boundary condition. Received vertical current meter profile time series came with some uncertainties: Occasionally the surface velocity is  slower than in lower layers which can occur due to transient dynamics such as wind forcing. The vertical profile spacing from the sea surface, with a top layer of \SI{1.2}{\meter} and other layers of \SI{.5}{\meter}, exhibited cumulative layer thicknesses that occasionally did not match the total measured depth. Therefore, a considerable uncertainty in the layer thickness had to be assumed and a sensitivity study conducted for the same. 

Equation \ref{eq:zid} can become ill-posed to such uncertainties which will not always average out: If, for example, there is only a minuscule difference between the two velocities in the denominator, then the roughness length is overestimated by logarithmic orders of magnitude. Likewise, the equation is sensitive to an error in height $z_1$. Consequentially, not the difference between adjacent \SI{.5}{\meter} thin layers but two layers apart, that is, layer $3$ and $5$, have been considered.\newline
The bottom layer, in principle, would have been more indicative but has been disregarded to obtain a lower relative error for the reference height. Additionally, measurements were excluded that did not exhibit a significant or positive vertical velocity difference $\delta u = u_1 - u_2$.\newline
In order to ascertain a well-posed response despite the uncertainty in reference height and the filtration of small $\delta u$, both has been varied and the parameter identification conducted with $2,000$ measurements. The roughness length distribution is shown in Figure \ref{fig:ZID}. Particularly compared to the considerable fluctuations in surface friction, Figure \ref{fig:ZID} retains well-posedness and returns, regardless of set minimal $\delta$ in velocity and assumed reference height, a roughness length on the order of $.2$ \SI{}{\meter}. Regardless of the two varied parameter, a roughness length on the order of \SI{0.2}{\meter} is obtained.

%An initial bed friction value from the previous regulatory reviewed study of the site \cite{noauthor_environmental_2019} of $0.032 m$ was applied.

%The bed resistance parameter was set to 0.032 m as per a previous study \ref{}. The Smagorinsky coefficient for the horizontal eddy viscosity was set to the common lower limit of 0.1 \ref{} given the additional eddy diffusion that will occur due to numerical diffusion \ref{}.

\FloatBarrier

\subsection{Validation}

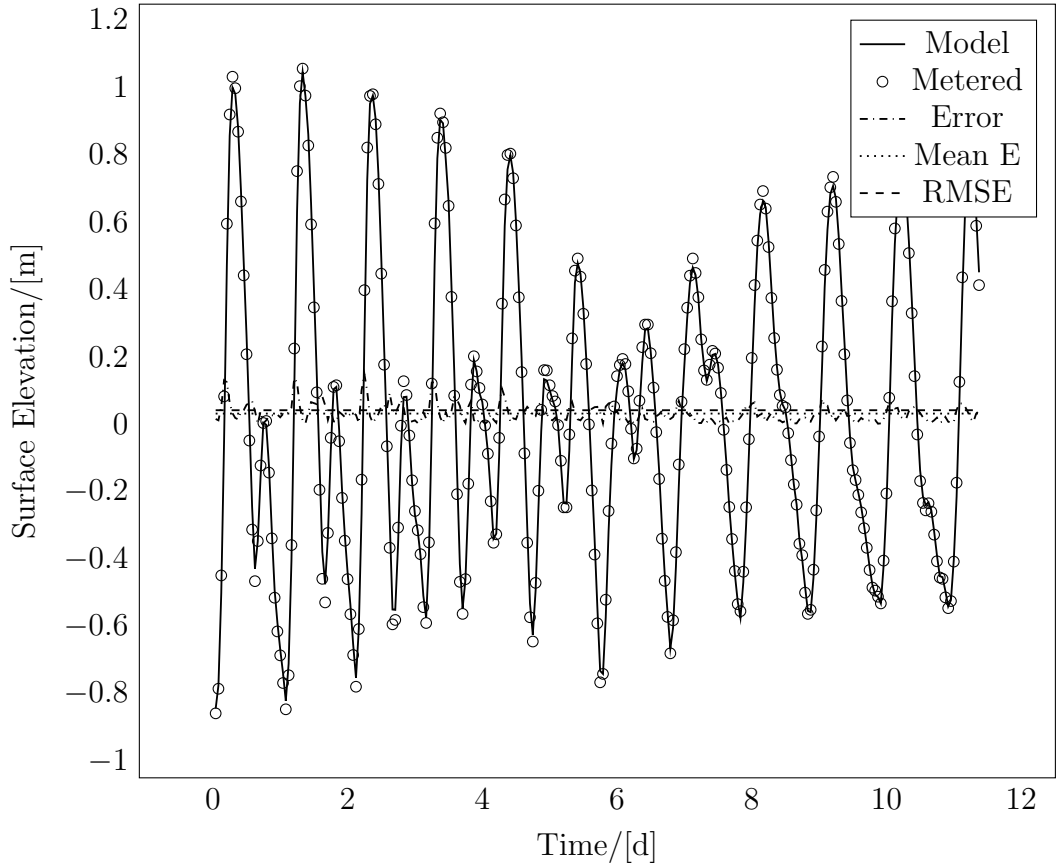
\begin{figure}[h!]
\centerline{
\begin{tikzpicture}[scale=1.0]
    \begin{axis}
    [   width=13.7cm,  % Adjust the width of the plot
    %height=6cm,  % Adjust the height of the plot
        xlabel={Time/[\si{\day}]},
        ylabel={Surface Elevation/[\si{\meter}]},
        axis line style={black}, 
            ]
            
        \addplot[line width=0.7pt, color=black] table[x=ti, y=si] {CM2sum.dat};       
        
        \addplot[only marks, mark=o, mark options={color=black}] table[x=ti, y=sm] {CM2sum.dat};             
        \addplot[dashdotted, line width=0.7pt,] table[x=ti, y=ea] {CM2sum.dat};             
        \addplot[dotted, line width=0.7pt,] table[x=ti, y=ma] {CM2sum.dat}; 

        \addplot[dashed, line width=0.7pt,] table[x=ti, y=rm] {CM2sum.dat};                         

        \legend{Model, Metered, Error, Mean E,RMSE}
        
    \end{axis}
\end{tikzpicture}
 }
\caption{Correlation of simulated and surveyed surface elevation at the location of current meter 2 which also recorded depth, that is, surface elevation in August 2023.}
\label{fig:CM2sum}
\end{figure}

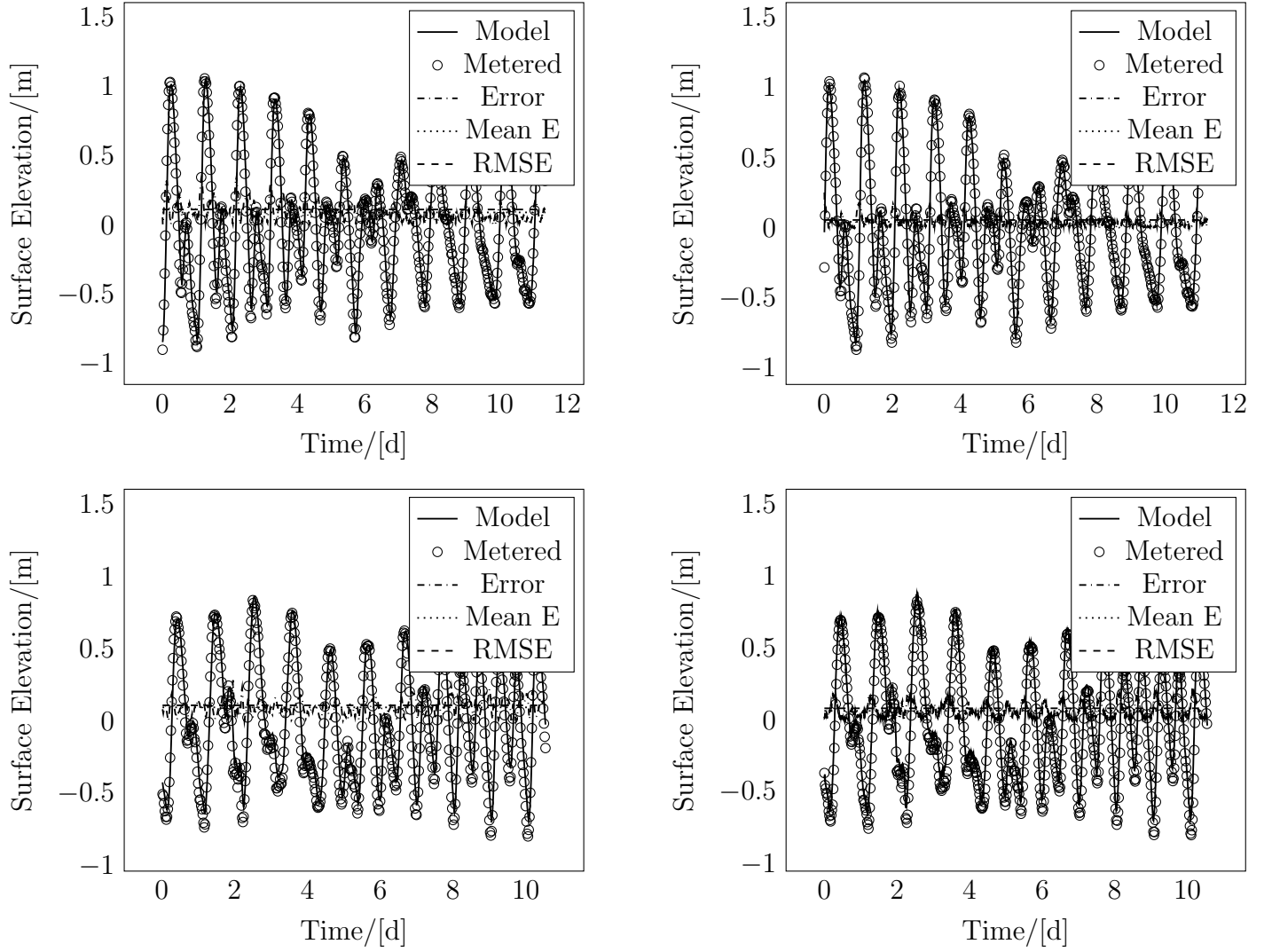
\begin{figure*}[h!]
\begin{adjustwidth}{-2cm}{-3cm}
  \begin{subfigure}[b]{0.7\textwidth}
\begin{tikzpicture}[scale=1.0]
    \begin{axis}
    [   
        xlabel={Time/[\si{\day}]},
        ylabel={Surface Elevation/[\si{\meter}]},
        axis line style={black}, 
        ymax=1.6,
            ]
        \addplot[line width=0.7pt, color=black] table[x=ti, y=si] {TDM2sum.dat};       
        
        \addplot[only marks, mark=o, mark options={color=black}, mark repeat=3] table[x=ti, y=sm] {TDM2sum.dat};             
        \addplot[dashdotted, line width=0.7pt,] table[x=ti, y=ea] {TDM2sum.dat};             
        \addplot[dotted, line width=0.7pt,] table[x=ti, y=ma] {TDM2sum.dat}; 

        \addplot[dashed, line width=0.7pt,] table[x=ti, y=rm] {TDM2sum.dat};                         

        \legend{Model, Metered, Error, Mean E,RMSE}
        
    \end{axis}
\end{tikzpicture}

%\label{fig:validitySUMMERT3}
\end{subfigure}
\hfill
  \begin{subfigure}[b]{0.7\textwidth}
\begin{tikzpicture}[scale=1.0]
    \begin{axis}
    [   
        xlabel={Time/[\si{\day}]},
        ylabel={Surface Elevation/[\si{\meter}]},
        axis line style={black}, 
        ymax=1.6,
            ]
        \addplot[line width=0.7pt, color=black] table[x=ti, y=si] {TDM3sum.dat};       
        
        \addplot[only marks, mark=o, mark options={color=black}, mark repeat=3] table[x=ti, y=sm] {TDM3sum.dat};             
        \addplot[dashdotted, line width=0.7pt,] table[x=ti, y=ea] {TDM3sum.dat};             
        \addplot[dotted, line width=0.7pt,] table[x=ti, y=ma] {TDM3sum.dat}; 

        \addplot[dashed, line width=0.7pt,] table[x=ti, y=rm] {TDM3sum.dat};                         

        \legend{Model, Metered, Error, Mean E,RMSE}
        
    \end{axis}
\end{tikzpicture}

%\label{fig:validitySUMMERT3}
\end{subfigure}

\medskip

  \begin{subfigure}[b]{0.7\textwidth}
\begin{tikzpicture}[scale=1.0]
    \begin{axis}
    [   
        xlabel={Time/[\si{\day}]},
        ylabel={Surface Elevation/[\si{\meter}]},
        axis line style={black}, 
        ymax=1.6,
            ]
        \addplot[line width=0.7pt, color=black] table[x=ti, y=si] {TDM2win.dat};       
        
        \addplot[only marks, mark=o, mark options={color=black}, mark repeat=1.5] table[x=ti, y=sm] {TDM2win.dat};             
        \addplot[dashdotted, line width=0.7pt] table[x=ti, y=ea] {TDM2win.dat};             
        \addplot[dotted, line width=0.7pt] table[x=ti, y=ma] {TDM2win.dat}; 

        \addplot[dashed, line width=0.7pt] table[x=ti, y=rm] {TDM2win.dat};                         

        \legend{Model, Metered, Error, Mean E,RMSE}
        
    \end{axis}
\end{tikzpicture}

%\label{fig:validitySUMMERT3}
\end{subfigure}
\hfill
  \begin{subfigure}[b]{0.7\textwidth}
\begin{tikzpicture}[scale=1.0]
    \begin{axis}
    [   
        xlabel={Time/[\si{\day}]},
        ylabel={Surface Elevation/[\si{\meter}]},
        axis line style={black}, 
        ymax=1.6,
            ]
        \addplot[line width=0.7pt, color=black] table[x=ti, y=si] {TDM3win.dat};       
        
        \addplot[only marks, mark=o, mark options={color=black}, mark repeat=5] table[x=ti, y=sm] {TDM3win.dat};             
        \addplot[dashdotted, line width=0.7pt,] table[x=ti, y=ea] {TDM3win.dat};             
        \addplot[dotted, line width=0.7pt,] table[x=ti, y=ma] {TDM3win.dat}; 

        \addplot[dashed, line width=0.7pt,] table[x=ti, y=rm] {TDM3win.dat};                         

        \legend{Model, Metered, Error, Mean E,RMSE}
        
    \end{axis}
\end{tikzpicture}
\end{subfigure}

\end{adjustwidth}
\caption{Correlation of simulated and surveyed surface elevation at the location of tidal meter 2 (left) and 3 (right) during August (top) and April (bottom) of 2023.}
\label{fig:validity}
\end{figure*}

Surface elevation predictions have been stored during the simulation for finite volumes that correspond to the tidal and current meter locations within the computational domain. Measured and simulated time series were vertically referenced to the observed mean sea level with \verb+TM = TM - mean(TM)+, correlated, and exhibited the accuracies compiled in Table {tab:validity} below. 
For the measured time series, the start time is specified with \verb+t1 = datetime(2022,9,14,12,0,0)+ and the end time translocated by \verb+hours(.5)*(length(TM)-1)+ with respect to the start time. The resultant time vector is build as per the data point frequency \verb+tTM=t1:hours(.5):t2+, reflective of the half-hourly sampling. The trivial \verb+hold on+ command provides for the superposition of the plots for the measured and simulated surface elevation with \verb+plot(tTM,TM)+ and so forth. \newline
Surveyed and simulated surface elevations as well as error, mean error and root mean square error (RMSE) are depicted in Figures \ref{fig:CM2sum} and \ref{fig:validity}. Table \ref{tab:validity} furthermore contains percent erros besides absolute errors, percent RMSE, as well as R\textsuperscript{2} and R-Pearson to quantify the quality of correlation.\newline
Two survey locations served to specify the boundary forcing and three survey locations served to validate the accuracy of the simulation. Given that two seasons have been examined, four time series were available for boundary forcing and six to validate the simulation.

%The commands for the validation are provided in the \hyperref[para:Valcommands]{appendix} in the proper order.

\begin{table}[h]
\centering
\caption{Validation of Simulated with Measured Surface Elevation. $*$ 0.9957 rounded after third digit.}
\begin{tabular}{c c c c c c c }
\arrayrulecolor{black}\hline
Quantity & CM2 & TM2 & TM3  & TM2&TM3\\[-2pt]
 & August & August & August & April& April \\
\hline
\SI{}{\percent} Error $\epsilon$ & 1.7 & 4.8 & 2.4&5.6 & 3.8 \\
Abs. $\epsilon$ & 3.3 & 9.4 & 4.6 & 8.9& 6.0\\[-2pt]
 $[\SI{}{\centi\meter}]$ &  &  &  & &\\
RMSE &  4.4 &  12& 5.9 & 10&7.6\\[-2pt]$[\SI{}{\centi\meter}]$ &  &  &  & &\\
\SI{}{\percent} NRMSE & 2.3 & 6.9 & 3.1 & 6.5&4.8\\
R\textsuperscript{2} & 0.99  & 0.94 & 0.98 & 0.94&0.97\\
R-Pearson & 1.0$^{*}$ & 0.97 & 0.99 & 0.97&0.99\\
\arrayrulecolor{black}\hline
\end{tabular}
\label{tab:validity}
\end{table}

%\newpage

% Load the data
%\pgfplotstableread{CM2sum.dat}\mydata

%
%\pgfplotstablecreatecol[
%    expr={\thisrow{ti}*2},
%]{newti}\mydata

%\pgfplotstableclear{\mydata}
% Multiply the x-values by three
%\pgfplotstablecreatecol[
%    expr={\thisrow{ti}*3}, % Multiply each value in the column "ti" by 3
%    colname=newti, % Name of the new column
%]{\mydata}

\FloatBarrier

 %The elevation distribution is shown for the entire domain and one time instance in Figure \ref{fig:flowfield}.

%\begin{figure*}[htbp] %b for bottom
%\includegraphics[width=1.0\linewidth]{flowfield.png}
%\caption{Flow field: water flowing from the south west to the north east as water is receding. The sea surface elevation transitions from levels in Doha bay to levels in the Arabian Gulf.}
%\label{fig:flowfield}
%\end{figure*}

\subsection{Waves, sediment entrainmnt, and settling}
\label{sec:waves}
 Wave motion has been resolved with a high resolution mesh for wave propagation with an even resolution throughout the entire domain. High energy waves as per the NOAA CFSR model have been compiled \cite{HRWallingford2017} for \SI{25.50}{\degree} N, \SI{52.17}{\degree} E, listed in Table \ref{tab:waves}, and wave transmission been modeled within Doha Bay \cite{TRUSTEngineering2023}. The latter found for the shortest and longest wavelength, $6.40$ and $7.12$ seconds, a transmission to significant wave heights on the order of $0.1$ and $0.4$ \SI{}{\meter} respectively. The same parameter were simulated with the wave-resolving model in order to resolve wave transmissions for Katara and account for groyne modifications. The periods of $6.4$ and $7.12$ \SI{}{\second} correspond to wavelengths $L = gT^2/(2\pi)$ of $64$ and $79$ \SI{}{\meter}, which were resolved with the  $\sim6$ \SI{}{\meter} fine high-resolution mesh.

\begin{table}[h]
\centering
\caption{High Energy Wave Conditions East of Safliya Island, NOAA CFSR}
\begin{tabular}{c c c c c }
\hline
Return period /[\SI{}{year}] & $H_s$ /[\SI{}{\meter}] & $T_p$ /[\SI{}{\second}] & Direction /[\SI{}{\degree}] \\
\hline
$100$ & 1.92 & 6.40 & 12.5 \\
$100$ & 2.34 & 7.12 & 57.5 \\
$100$ & 2.67 & 6.99 & 102.5 \\
$100$ & 2.05 & 5.72 & 147.5 \\
\hline
\end{tabular}
\label{tab:waves}
\end{table}

\begin{figure*}[h] %b for bottom
\centering
\begin{adjustbox}{center}
\pgfplotsset{compat=1.17, width=1.0*\linewidth}
\begin{tikzpicture}
    \begin{axis}[
        %axis equal image,
    xlabel={Eastern UTM/[10 \si{\meter}]}, % Using siunitx for unit
    ylabel={Northern UTM/[\si{\meter}]},
        xmin=552926.5,
        xmax=554686.5, 
        ymin=2804040.0,
        ymax=2805800,
        xtick={553000, 553200, 553400, 553600, 553800, 554000, 554200, 554400, 554600 }, 
        xticklabels={$55{,}300$, $55{,}320$, $55{,}340$, $55{,}360$, $55{,}380$, $55{,}400$, $55{,}420$, $55{,}440$, $55{,}460$},
        ytick={2804200, 2804400, 2804600, 2804800, 2805000, 2805200, 2805400, 2805600, 2805800}, 
        yticklabels={$2{,}804{,}200$, $2{,}804{,}400$, $2{,}804{,}600$, $2{,}804{,}800$, $2{,}805{,}000$, $2{,}805{,}200$, $2{,}805{,}400$, $2{,}805{,}600$, $2{,}805{,}800$},
        tick style={draw=none}, 
        axis line style={draw=none},
        width=0.965*\linewidth,
        height=0.965*\linewidth,
        scaled ticks=false 
    ]   

    \addplot graphics [        xmin=552926.5,
        xmax=554686.5, 
        ymin=2804040.0,
        ymax=2805800] {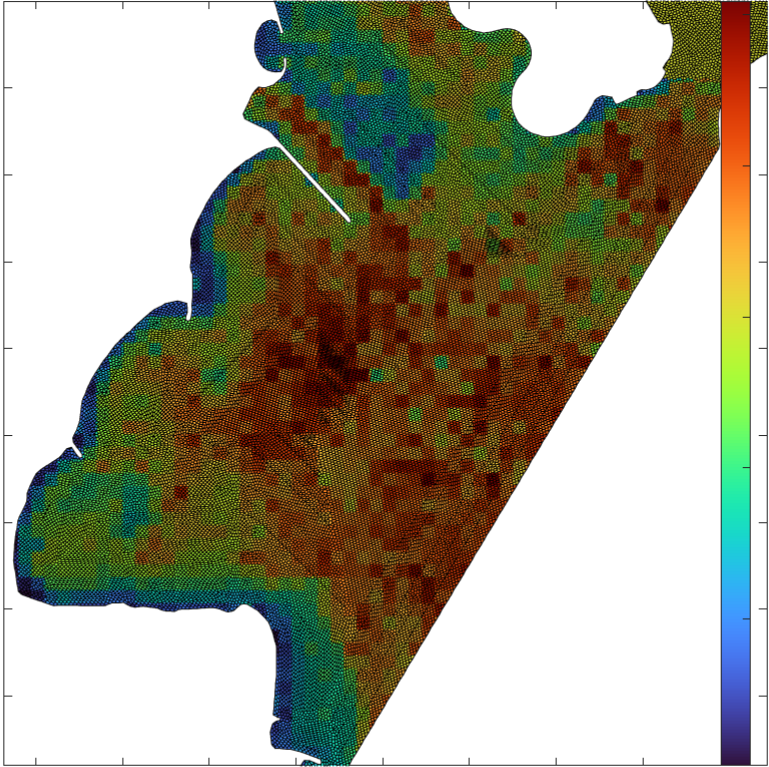};
        
    \end{axis}  
    \begin{axis}[ylabel={Depth/[\si{\meter}]}, 
    hide x axis, %as we just want right y axis for color bar
    %axis equal image,
    major tick length=0pt, % Make major ticks invisible
    minor tick length=0pt, % Make minor ticks invisible
    xmin=0, xmax=5.09, %set to value that puts right beside box 
            ytick={1, 2, 3, 4, 5},
        yticklabels={$1$, $2$, $3$, $4$, $5$},
                tick style={draw=none}, 
        axis line style={draw=none},
    ymin=0.03, ymax=5.09, clip=false, axis y line*=right,  ylabel near ticks,
            width=0.965*\linewidth,
        height=0.965*\linewidth
    ]
    \end{axis}
    \end{tikzpicture}
    \end{adjustbox}
    \caption{Wave-resolving high resolution mesh to bring into focus the propagation of high energy waves. The northern portion of the mesh is not shown to resolve individual cells in the plot.}
\label{fig:wavemesh}
\end{figure*}

\begin{figure*}[h] %b for bottom
\centering
\begin{adjustbox}{center}
\pgfplotsset{compat=1.17, width=1.0*\linewidth}
\begin{tikzpicture}
    \begin{axis}[
    xlabel={Eastern UTM/[\si{\meter}]},
    ylabel={Northern UTM/[\si{\meter}]},
        xmin=552383.5,
        xmax=555698.6, 
        ymin=2803880,
        ymax=2806500,
        xtick={552500,553000, 553500, 554000, 554500, 555000, 555500 }, 
        xticklabels={$552{,}500$, $553{,}000$, $553{,}500$, $554{,}000$, $554{,}500$, $555{,}000$, $555{,}500$},
        ytick={2804000, 2804500, 2805000, 2805500, 2806000}, 
        yticklabels={$2{,}804{,}000$, $2{,}804{,}500$, $2{,}805{,}000$, $2{,}805{,}500$, $2{,}806{,}000$},
        tick style={draw=none}, 
        axis line style={draw=none},
        width=1*\linewidth,
        height=.79*\linewidth,
        scaled ticks=false,
    ]   

    \addplot graphics [               xmin=552383.5,
        xmax=555698.6,
        ymin=2803880,
        ymax=2806500] {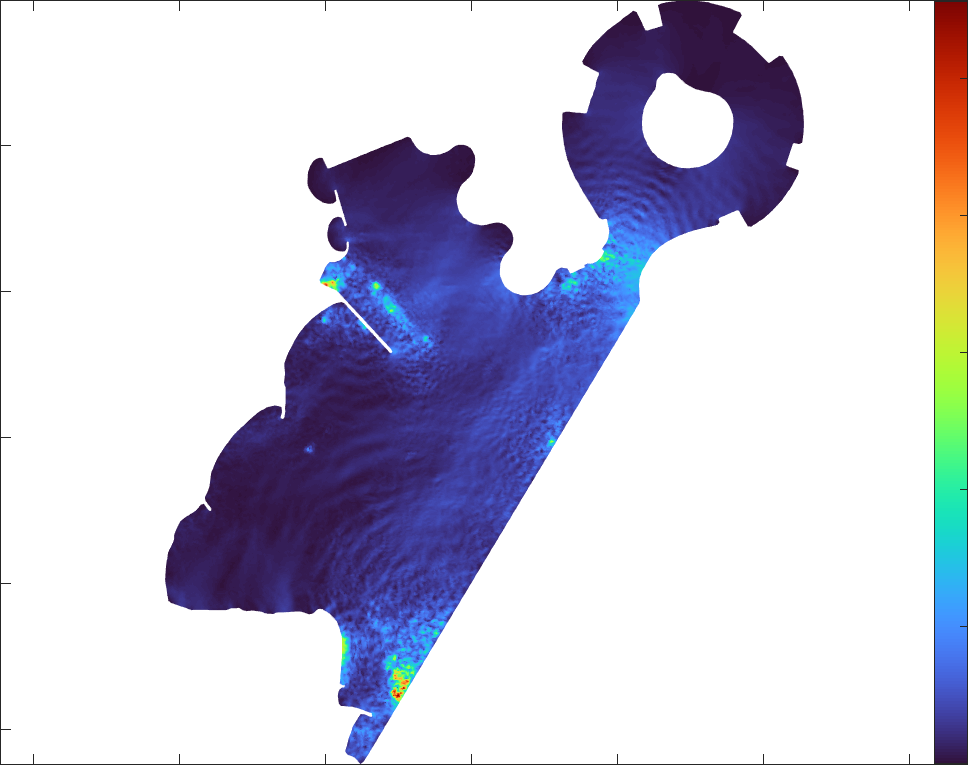};
        
    \end{axis}  
    \begin{axis}[ylabel={Velocity Magnitude/[\si{\meter/\second}]}, 
    hide x axis,
    major tick length=0pt,
    minor tick length=0pt, 
    xmin=0, xmax=0.2785, 
    ytick={0.05, 0.1, 0.15, 0.2, 0.25},
    yticklabels={$.05$, $.1$, $.15$, $.2$, $.25$},
    tick style={draw=none}, 
    axis line style={draw=none},
    ymin=0.0, ymax=0.2785, clip=false, axis y line*=right,  ylabel near ticks,
    width=1*\linewidth,
    height=0.79*\linewidth
    ]
    \end{axis}
    \end{tikzpicture}
    \end{adjustbox}
    \caption{Wave-resolving simulation of tidal regime, showing the velocity
magnitude distribution.}
\label{fig:tidalwaves}
\end{figure*}

\begin{figure*}[h] %b for bottom
\centering
\begin{adjustbox}{center}
\pgfplotsset{compat=1.17, width=1.0*\linewidth}
\begin{tikzpicture}
    \begin{axis}[
    xlabel={Eastern UTM/[\si{\meter}]},
    ylabel={Northern UTM/[\si{\meter}]},
        xmin=552383.5,
        xmax=555698.6, 
        ymin=2803880,
        ymax=2806500,
        xtick={552500,553000, 553500, 554000, 554500, 555000, 555500 }, 
        xticklabels={$552{,}500$, $553{,}000$, $553{,}500$, $554{,}000$, $554{,}500$, $555{,}000$, $555{,}500$},
        ytick={2804000, 2804500, 2805000, 2805500, 2806000}, 
        yticklabels={$2{,}804{,}000$, $2{,}804{,}500$, $2{,}805{,}000$, $2{,}805{,}500$, $2{,}806{,}000$},
        tick style={draw=none}, 
        axis line style={draw=none},
        width=1*\linewidth,
        height=.79*\linewidth,
        scaled ticks=false,
    ]   

    \addplot graphics [               xmin=552383.5,
        xmax=555698.6,
        ymin=2803880,
        ymax=2806500] {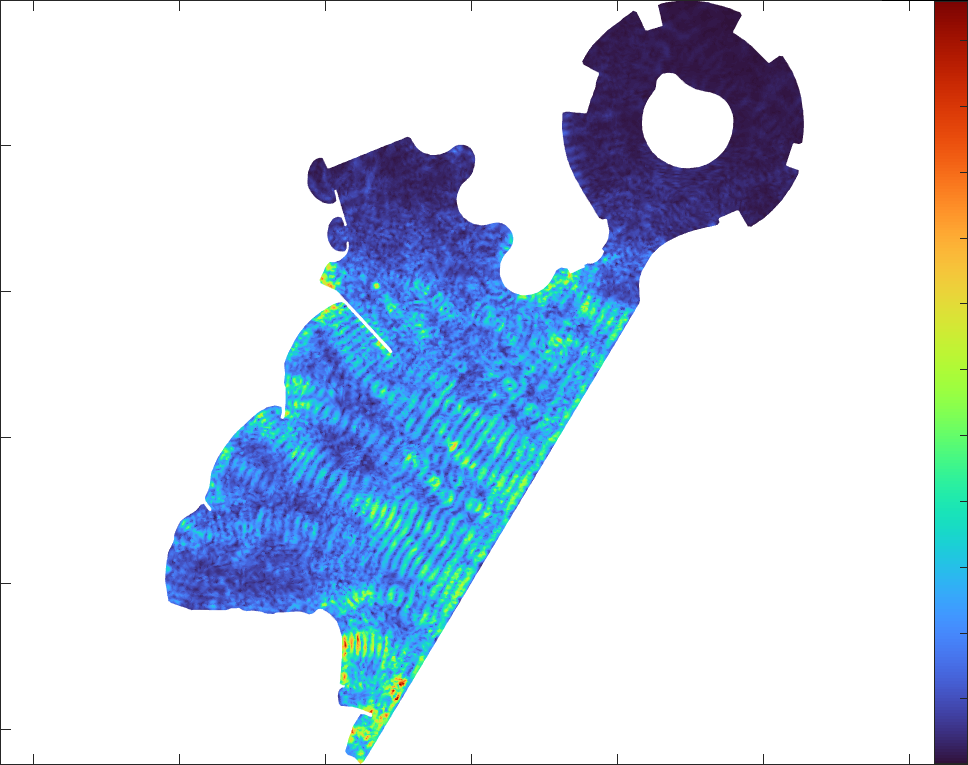};
        
    \end{axis}  
    \begin{axis}[ylabel={Friction Velocity/[\si{\meter/\second}]}, 
    hide x axis,
    major tick length=0pt,
    minor tick length=0pt, 
    xmin=0, xmax=5.8, 
    ytick={.5, 1, 1.5, 2, 2.5, 3, 3.5, 4, 4.5, 5, 5.5},
    yticklabels={$.005$, $.01$, $.015$, $.02$, $.025$, $.03$, $.035$, $.04$, $.045$, $.05$, $.055$},
    tick style={draw=none}, 
    axis line style={draw=none},
    ymin=0.0, ymax=5.8, clip=false, axis y line*=right,  ylabel near ticks,
    width=1*\linewidth,
    height=0.79*\linewidth
    ]
    \end{axis}
    \end{tikzpicture}
    \end{adjustbox}
    \caption{Wave-resolving simulation of high energy regime, showing a moment in time for the dynamic bottom friction velocity distribution. The latter governs erosion.}
\label{fig:HEwaves}
\end{figure*}

Fine structures, shown in Figure \ref{fig:tidalwaves}, in tidal currents have also been resolved with the high resolution mesh.
Both, horizontal geometry and bathymetry, determine the transformation of incoming waves. Acceleration due to continuity at narrow or shallow sections yield an acceleration in both, the depth-averaged and the friction velocity, the latter driving the entrainment of sediment and, hence, erosion. \newline
Second order Stokes waves enter the high resolution domain as depicted for the friction velocity in Figure \ref{fig:HEwaves}. The boundary of the high resolution domain is aligned with the wave direction \cite{TRUSTEngineering2023} with the waves being superimposed to the tidal boundary condition. High friction velocity entails erosion. The dynamic friction velocity distribution can, thus, already reveal spots that are vulnerable to morphological changes. \newline
The dynamic Rouse number distribution visualizes where sediment settling and erosion predominate. Values below and above 1 correspond to erosion and settling respectively. Island developments constitute a perturbation of the natural coastal ocean equilibrium in sediment transport. The highly resolved simulation, shown in Figure \ref{fig:erosionratio}, brings into focus fine pattern and structures, enhancing the reliability of on the former based mitigating measures for artificial coastal geometries. 
\newline
The Rouse number distribution was simulated, resolving wavelengths on the order of 60 \SI{}{\meter}, on a 6 \SI{}{\meter} Voronoi mesh, avoiding wave fronts on acute finite volume polygon angles. The tile pattern exhibited in Figure \ref{fig:wavemesh} stems from the $30$ \SI{}{\meter} resolution of the open source Landsat satellite images utilized in remote sensing. The mesh has, thus, a higher resolution than the bathymetric model. Nevertheless, commercially available satellite images can faciliate a remote sensing resolution that can match the resolution of the mesh.

A high resolution mesh was produced for Katara beaches with the wave boundary being aligned with the wave direction. Wave resolving currents were simulated for sheltered and high energy conditions and the Rouse number distribution shown for the latter. If the wave’s orbital motion does not reach the seafloor, then perturbations of bottom layer currents are small. But waves with wavelengths in excess of the water depth do exert an influence on bottom currents with the latter governing shear and, thus, erosion. Such medium and long wavelength waves can result due to tidal forcing, displayed in Figure \ref{fig:tidalwaves}, as seiches and long wave agitation. 

The friction velocity distribution, displayed in Figure \ref{fig:HEwaves}, and ratio between settling and erosion fluxes were visualized with the Rouse number, displayed in Figure \ref{fig:erosionratio}, recovering prior observed phenomena: erosion at the southernmost Katara beach and sediment settling at the sheltered beach in the south-west. Sediment settling is found in both, the model and observations, to occur in the sheltered southernmost of the three crescent beaches. In prior studies of Doha Bay unresolved scales were broght into focus, revealing erosion being impeded adjacent to recently added groynes. However, for the northern groyne, erosion is impeded only south of it due to the angle of the groyne. 
\newline
That is, the wave-resolving simulation brought fine patterns into focus that might not be recovered absent the deployed high resolution, enhancing coastal management. This encourages to conduct sediment transport simulations on Voronoi mesh-based platforms and with high resolution enabled by parallelization or GPU acceleration. \newline The wave-resolving simulation, displayed in Figure \ref{fig:HEwaves}, resolves wave-driven dynamics that dominate the friction velocity and wave attenuation in the Marina north of the development's beaches. 

\section{Conclusion}
Simulated surface elevations have been been validated with five time series from three tidal meters and for two seasons, April and Augsut 2023. The model exceeds real-time performance on a Ryzen 9 or comparable desktop CPU. Vertical current profile data were used to calibrate and conduct a sensitivity study for the roughness length, boundary conditions were set based on two tidal meter, and the validation conducted based on data from three locations, totalling five locations for validation, calibration, and boundary conditions.\newline
The correlation of simulated and surveyed surface elevation time series, exhibited an exceptionally precise correlation, data and simulation being for some plots visually identical, yielding a high confidence in model results. Mean error and RMSE were consistently below 7 \SI{}{\percent} as visualized in Figures \ref{fig:CM2sum} and \ref{fig:validity} as well as compiled in Table \ref{tab:validity}.
\newline
With Voronoi-capable models a reduction in cell count, numerical diffusion \cite{Holleman2013, chansolution2018}, and, as demonstrated herein, acute polygon angles can be achieved.\newline 
The model structure aligns with seamless pre- and post-processing in Matlab as documented in the paper, automatic parallelization, and seamless GPU acceleration. Documented are different approximations for some terms of the NS PDE, listed in Table \ref{tab:fveapprox}. 

\begin{figure*}[h] %b for bottom
\centering
\begin{adjustbox}{center}
\pgfplotsset{compat=1.17, width=1.0*\linewidth}
\begin{tikzpicture}
    \begin{axis}[
    xlabel={Eastern UTM/[\si{\meter}]},
    ylabel={Northern UTM/[\si{\meter}]},
        xmin=552575.5,
        xmax=555507.2, 
        ymin=2803870.5,
        ymax=2806507.2,
        xtick={553000, 553500, 554000, 554500, 555000}, 
        xticklabels={ $553{,}000$, $553{,}500$, $554{,}000$, $554{,}500$, $555{,}000$},
        ytick={2804000, 2804500, 2805000, 2805500, 2806000, 2806500}, 
        yticklabels={$2{,}804{,}000$, $2{,}804{,}500$, $2{,}805{,}000$, $2{,}805{,}500$, $2{,}806{,}000$,
        $2{,}806{,}500$},
        tick style={draw=none}, 
        axis line style={draw=none},
        width=1*\linewidth,
        height=.8995*\linewidth,
        scaled ticks=false,
    ]   

    \addplot graphics [               xmin=552575.5,
        xmax=555507.2,
        ymin=2803870.5,
        ymax=2806507.2] {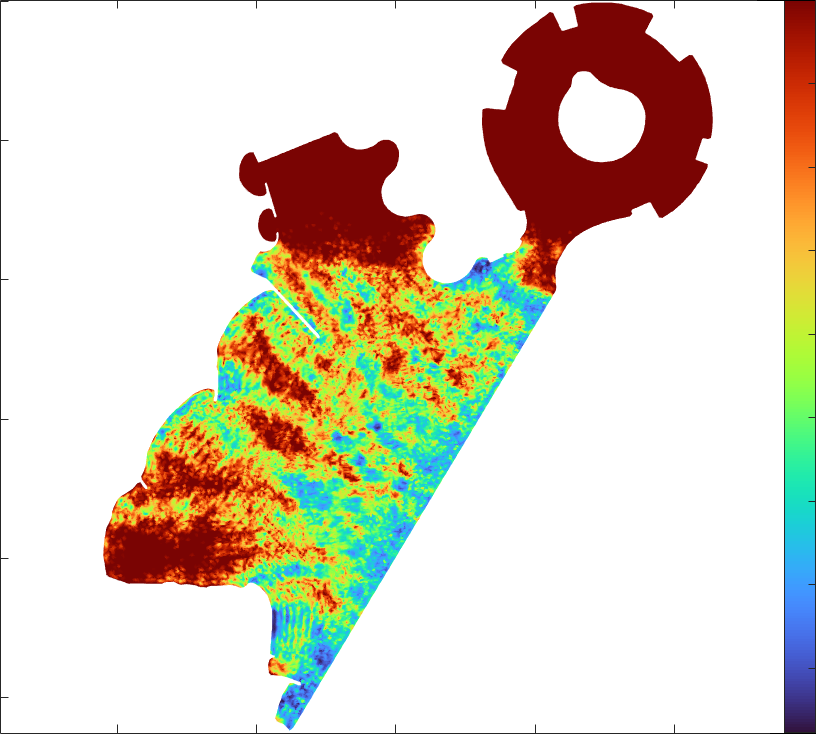};
        
    \end{axis}  
    \begin{axis}[ylabel={Ratio of Settling and Erosive Flux}, 
    hide x axis,
    major tick length=0pt,
    minor tick length=0pt, 
    xmin=0.24, xmax=2, 
    ytick={.4, 0.6, .8, 1, 1.2, 1.4, 1.6, 1.8, 2},
    yticklabels={$.4$, $.6$, $.8$, $1$, $1.2$, $.4$, $1.6$, $1.8$, $2$},
    tick style={draw=none}, 
    axis line style={draw=none},
    ymin=0.24, ymax=2, clip=false, axis y line*=right,  ylabel near ticks,
    width=1*\linewidth,
    height=0.8985*\linewidth
    ]
    \end{axis}
    \end{tikzpicture}
    \end{adjustbox}
    \caption{Ratio of sediment settling and erosion fluxes, showing  sediment settling at the southern of the three Katara beaches (red), erosion (blue) at the southernmost beach and the functioning of the groynes. The northern groyne that is tilted to the south only shelters its southern side.}
\label{fig:erosionratio}
\end{figure*}

These alternatives additionally provide for cross-correlation between different solvers, readily providing a discrepancy-based error estimate for adaptive time stepping.\newline 
The model comes with a comprehensive environment of modules, a remote sensing module with spiking neuron filtration, published prior \cite{w14050810}, its pollutant fate transport model for nonlinear conversion, published prior \cite{LAWEN2013330, Lawen2014}, and the Voronoi mesh generator published separately.\newline
The simulated, highly resolved dynamic Rouse number distribution, the ratio between sediment settling and erosive flux, displayed in Figure \ref{fig:erosionratio}, includes orbital wave motion. Otherwise unresolved details in settling and erosion, particularly adjacent to the groynes, have been recovered in Figure \ref{fig:erosionratio}. Wave-resolved simulations can, therefore, considerably enhance coastal management.\newline
Voronoi schemes can be expanded to n dimensions. For coastal systems that might not improve results: the usual approach \cite{LAWEN201099, LAWEN2013330, Lawen2014} of resolving the vertical rather via multiple layers retains an alignment with the dominant horizontal current components and, thus, avoids numerical diffusion. That is, retaining multiple layers achieves quasi flow alignment for the vertical. This caution might not hold for modeling wave breaking or moving coastal meshes (4D Voronoi). For comprehensiveness, the development of a global model may be a subsequent stage, a step that likewise provides boundary conditions for regional and local models \cite{CouplingofSeaLevelRiseTidalAmplificationandInundation, 10.1016/j.cageo.2015.08.010}.

    % Add axis for cbar
 %   \begin{axis}[ylabel=Depth in meters, 
 %   hide x axis, %as we just want right y axis for color bar
 %   axis equal image,
 %   major tick length=0pt, % Make major ticks invisible
 %   minor tick length=0pt, % Make minor ticks invisible
 %   xmin=0, xmax=5.25, %set to value that puts right beside box 
 %   ymin=0.03, ymax=5.09, clip=false, axis y line*=right,  ylabel near ticks 
 %   ]
 %   \end{axis}

\newpage
\begin{appendices}
\section{Bathymetry Survey Coverage}

\begin{figure*}[htbp] %b for bottom
\centering
\pgfplotsset{compat=1.17, width=1.0*\linewidth
}
\begin{tikzpicture}
    \begin{axis}[
        axis equal image,
        xlabel={Eastern QND},
        ylabel={Northern QND},
        xmin=230942,
        xmax=232369,
        ymin=399702,
        ymax=401202,
        xtick={231000, 231200, 231400, 231600, 231800, 232000, 232200}, % Set the tick locations on the x-axis
%        x tick label style={
%            /pgf/number format/.cd,
%            use comma,
%            fixed,
%            fixed zerofill,
%            precision=0,
%            1000 sep={}
%        },
        xticklabels={$231{,}000$, $231{,}200$, $231{,}400$, $231{,}600$, $231{,}800$, $232{,}000$, $232{,}200$}, % Set the labels for the x-axis ticks
        ytick={399800, 400000, 400200, 400400, 400600, 400800, 401000}, % Set the tick locations on the y-axis
        yticklabels={$399{,}800$, $400{,}000$, $400{,}200$, $400{,}400$, $400{,}600$, $400{,}800$, $401{,}000$},
        %width=1.0*\linewidth, 
        scaled ticks=false 
    ]
    \node[anchor=south west,inner sep=0] at (230942,399702) {
    \includegraphics[width=0.739\linewidth]{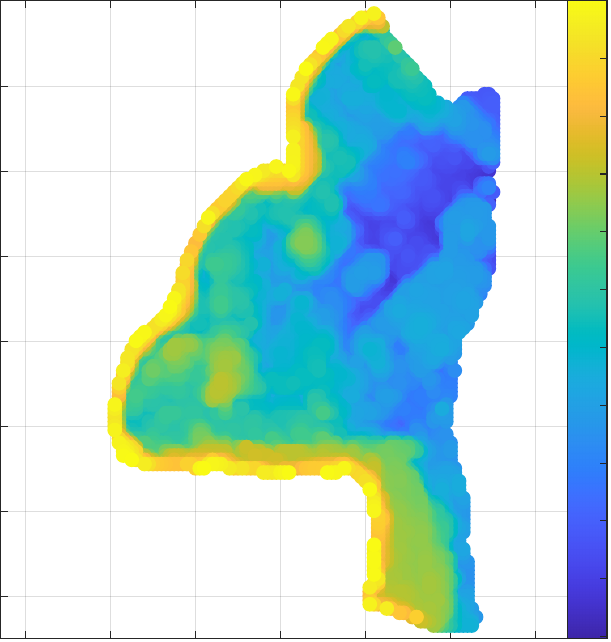}}; 
    \end{axis}
    
    % Add axis for cbar
    \begin{axis}[ylabel=Depth in meters, 
    hide x axis, %as we just want right y axis for color bar
    axis equal image,
    major tick length=0pt, % Make major ticks invisible
    minor tick length=0pt, % Make minor ticks invisible
    xmin=0, xmax=5.25, %set to value that puts right beside box 
    ymin=-5.515, ymax=0.002, clip=false, axis y line*=right,  ylabel near ticks 
    ]
    \end{axis}
\end{tikzpicture}
\caption{Bathymetry measurements during survey off Katara beaches. Not covered areas where augmented with remote sensing published separately \cite{w14050810}.}
\label{fig:newmesh}
\end{figure*}
\end{appendices}

%\printbibliography
\bibliographystyle{unsrt}
\bibliography{references}  %%% Uncomment this line and comment out the ``thebibliography'' section below to use the external .bib file (using bibtex) .





% \end{thebibliography}

\end{document}